\documentclass[aps, pra,  a4paper, showpacs, twocolumn, english, 10pt]{revtex4-2}
\usepackage{bbm, amsthm, bm, textcomp, nicefrac, geometry,ragged2e}
\usepackage[dvipsnames]{xcolor}
\usepackage{float}
\usepackage[bbgreekl]{mathbbol}
\usepackage{graphicx,epstopdf,verbatim,enumitem}
\usepackage{pbox,array}
\usepackage{mathrsfs}
\usepackage{amssymb}
\usepackage{textgreek}
\usepackage{mathtools}
\usepackage{stackrel}
\usepackage[thinlines]{easytable}
\usepackage{amsmath}
\usepackage[utf8]{inputenc}
\usepackage{booktabs}
\usepackage{array}
\makeatletter
\usepackage{soul}
\usepackage[caption=false]{subfig}
\usepackage[dvipsnames]{xcolor}
\usepackage{hyperref}
\hypersetup{
    colorlinks = true,
    linkcolor = teal,
    citecolor = teal,
    filecolor = teal,      
    urlcolor = teal,
}

\usepackage[T1]{fontenc}
\usepackage[caption=false]{subfig}
\usepackage{babel}
\usepackage[linesnumbered,ruled]{algorithm2e}
\usepackage[normalem]{ulem}

\addto\captionsenglish{}

\newcommand\id{\mathbbm{1}}
\newcommand{\ketbra}[2]{\left| #1\right\rangle\!\left\langle#2\right|}
\newcommand{\bo}[1]{\boldsymbol{#1}}

\newcommand{\ket}[1]{\left|#1\right\rangle}

\newcommand{\ti}{\text{i}}

\newcommand{\bra}[1]{\left\langle #1\right|}
\newcommand{\bracket}[2]{\left\langle #1|#2\right\rangle}
\newcommand{\proj}[1]{\ket{#1}\!\bra{#1}}

\definecolor{blockred}{rgb}{0.8, 0.0, 0.0}

\begin{document}

\title{Quantum Simulation of Noisy Quantum Networks}

\author{Ferran Riera-S\`abat}
\thanks{These two authors contributed equally. } 

\author{Jorge Miguel-Ramiro}
\thanks{These two authors contributed equally. }

\author{Wolfgang D\"ur}
\affiliation{Universit\"at Innsbruck, Institut f\"ur Theoretische Physik, Technikerstra{\ss}e 21a, Innsbruck 6020, Austria}

\date{\today}
\begin{abstract}
Complex quantum networks are not only hard to establish, but also difficult to simulate due to the exponentially growing state space and noise-induced imperfections. In this work, we propose an alternative approach that leverage quantum computers and noisy intermediate-scale quantum (NISQ) devices as simulators for quantum networks, including noisy quantum devices, channels, and protocols. Rather than considering noise as an undesired property that needs to be mitigated, we demonstrate how imperfections in quantum hardware can be utilized to simulate real-world communication devices under realistic conditions beyond classical simulation capabilities. Our approach allows NISQ devices with modest noise to simulate devices with more significant imperfections enabling large-scale, detailed simulations of quantum networks, where exact error models can be treated.  It also improves over direct implementation and benchmarking of real networks, as waiting times for information transmission, locality, and memory restrictions do not apply. This framework can offer advantages in flexibility, scalability, and precision, demonstrating that NISQ devices can serve as natural testbeds for complex quantum networks, and paving the way for more efficient quantum network simulations.


\end{abstract}

\maketitle

\tableofcontents

\section{Introduction}
Quantum technologies utilize the features of quantum systems and offer the possibility for powerful and unexplored applications \cite{Preskill2018, Grumbling2019, Petschnigg2019, Bayer2021, Cho_2021}. Quantum computers promise to solve optimization and other problems in logistics \cite{Hassija2020, Hernandez2020}, finances \cite{Ors2019, Herman2022}, chemistry \cite{McArdle2020, Cao2019} and drug design \cite{Cao2018} that are not accessible with classical devices. Quantum simulators \cite{Buluta2009, Cirac2012, Franco14, Bassoli2020, Bhowmick2023} are discussed as tools to simulate physical models and provide insights into phenomena such as high-temperature superconductivity \cite{Zhou2021}, and quantum sensors \cite{Degen2017} offer the possibility to measure physical quantities with unprecedented precision. Connecting such quantum devices to form a quantum network \cite{Kozlowski2019,Azuma2021}, or ultimately a quantum internet \cite{Kimble2008, Wehner2018internet}, leads to even more possibilities and unleashes the full power of quantum devices. Connecting quantum sensors to a quantum sensor network \cite{Eldredge2018,Sekatski2020} opens the way to measure spatially correlated quantities and applications such as high-resolution imaging. Connecting small-scale quantum computers makes them even more powerful, allowing access to the exponentially large Hilbert space. Planning and building such quantum networks is hence of importance not only to connect quantum devices and make them more powerful but also to make them broadly accessible. 

However, the same features that are responsible for the power of connected quantum devices also pose the main hindrance to simulating these systems. The exponentially growing state space makes an exact and accurate description of devices and protocols very demanding, if not impossible. Nevertheless, classical quantum network simulators  \cite{Coopmans_2021,Wu_2021,Satoh_2022,Julius2022, Chen_2023, Bel2024} have been developed and have been applied to simulate certain network devices and protocols, including a quantum switch  \cite{Dai2021}, repeater protocols \cite{Behera2019,Das2021}, satellite-based communication \cite{Wallnfer2022} or inter-network protocols \cite{Behera2019, Wenbo2022, Thotakura2023}. In order to cope with the inherent difficulties, the treated systems are either significantly small in size (at most a few tens of qubits) or very simplified models that are used to describe quantum states and noise processes. Techniques such as the stabilizer formalism \cite{Gottesman1997,GottesmanThesis,Aaronson2004,Nest2004} allow one to treat specific cases of interest efficiently, even for large systems, i.e., networks of multiple nodes, however when dealing with general situations, in particular noise that is not described by a simple error model such as local dephasing or depolarization, the effort to perform an exact simulation scales exponentially with the number of systems, i.e., the number of nodes in the network, limiting the size of simulations based on classical means.

In this work, we explore how to use noisy intermediate-scale quantum (NISQ) devices \cite{Preskill2018} to perform the simulation of quantum networks and their features, see Fig.~\ref{fig:basicidea}.  The central idea of our work is both simple and natural: rather than relying on classical simulations, which make simulation of large processes or complex noises infeasible, we propose using quantum computers themselves to simulate these systems. This not only overcomes the problem of exponentially growing state space, but has the additional bonus that noise in the NISQ devices does not pose an obstacle for successful simulation, but is rather a feature that we utilize. Usually in applications of quantum technologies noise and imperfections are undesired and hinder us from harnessing the full power of quantum devices, and hence need to be suppressed and mitigated. 

Here, however, we aim to simulate quantum devices that are themselves noisy, and their interaction via even noisier quantum channels -- and it is exactly the performance of the quantum network taking such noise and imperfections into account that we are interested in. Our Ansatz is hence to modify and tailor the noise inherent in quantum computers in such a way that it represents the one of the device, gate or channel to be simulated. We consider relevant error models to describe noisy devices such as sources, channels, and memories, including dephasing, depolarizing, and Pauli channels, but also amplitude damping, absorption or combinations thereof. To this aim we introduce techniques and optimizations of how to simulate noise processes and gates taking imperfections into account - not by considering an idealized, perfect implementation and studying the effect of noise, but by actively reshaping the existing noise to obtain a highly accurate approximation of the desired target noise model for storage, gates, and channels. In many relevant cases, even exact modeling of the desired target noise model is possible, e.g., when dealing with Pauli noise channels in the NISQ devices with errors that are smaller than the one of the target system. Notice that noise in channels and memories in quantum communication networks is typically much larger than noise in local devices or quantum computers, and the local devices in network nodes may simply consist of small-scale quantum computers themselves. 

Quantum simulation offers multiple advantages over both classical simulation methods \cite{Coopmans_2021,Wu_2021,Satoh_2022,Julius2022, Chen_2023, Bel2024} and direct experimental benchmarking of quantum networks \cite{Liao2022,Helsen2023}. Classical simulators, despite providing valuable insights into the design and testing of quantum protocols, are inherently limited by the exponential growth of the state space, especially when simulating general noise models or large-scale topologies. To stay tractable, they often rely on strong simplifications or small system sizes, restricting their predictive power for realistic networks. Experimental benchmarking, on the other hand, while providing accurate information about real-world processes, is time-consuming, resource-intensive, and constrained by non-local current hardware capabilities and experimental setups.

In contrast, quantum simulation using NISQ devices combines flexibility with scalability \cite{Iten2017, qNetVO, Doolittle2023}. It allows to emulate realistic or even worst-case noise conditions, explore a broad parameter space, and study large or future network configurations that are beyond current experimental reach. Unlike benchmarking, it enables rapid characterization, protocol optimization, and feasibility assessments with minimal overhead. Moreover, quantum simulation avoids limitations from probabilistic processes, limited measurement access, or classical communication latency, which are intrinsic to real-world network tests.

We explore the central challenges involved in making quantum hardware a viable tool for simulating quantum networks. The main idea is to reshape the native noise present in NISQ processors so that it closely mimics the target noise of realistic components in quantum networks. This enables the high-fidelity simulation of quantum memories, gates, and channels, effectively turning hardware imperfections from a limitation into an  advantage.


The paper is organized as follows. In Sec.~\ref{sec:comparison} we qualitatively explain the details of our general idea for simulating quantum networks in quantum devices, by comparing our approach with other classical simulation techniques, as well as with benchmarking of real networks. In Sec.~\ref{sec:perfect_devices} we introduce basic tools and strategies that evidence the benefits of our approaches.  While in Sec.~\ref{sec:approaches}, we analyze different ways to simulate noise processes using unitary operations, ancilla qubits, measurements, and stochastic processes, in Sec.~\ref{sec:noisy_devices} we consider NISQ devices and discuss different ways to obtain target noise channels and processes, concentrating on single- and two-qubit maps. We treat different models to describe the simulating device, where we consider a direct description of the imperfect completely positive map and of imperfect unitary gates, where we demonstrate that utilizing and actively shaping noise outperforms a direct approach where simulation schemes developed for noiseless devices are implemented on NISQ devices. We use a variational black-box ansatz to find optimized implementations of specific target noise maps. We summarize and conclude in Sec.~\ref{sec:conclusions}.


\section{Advantages of quantum simulation of quantum networks} \label{sec:comparison}

\begin{figure}
    \centering
    \includegraphics[width=\columnwidth]{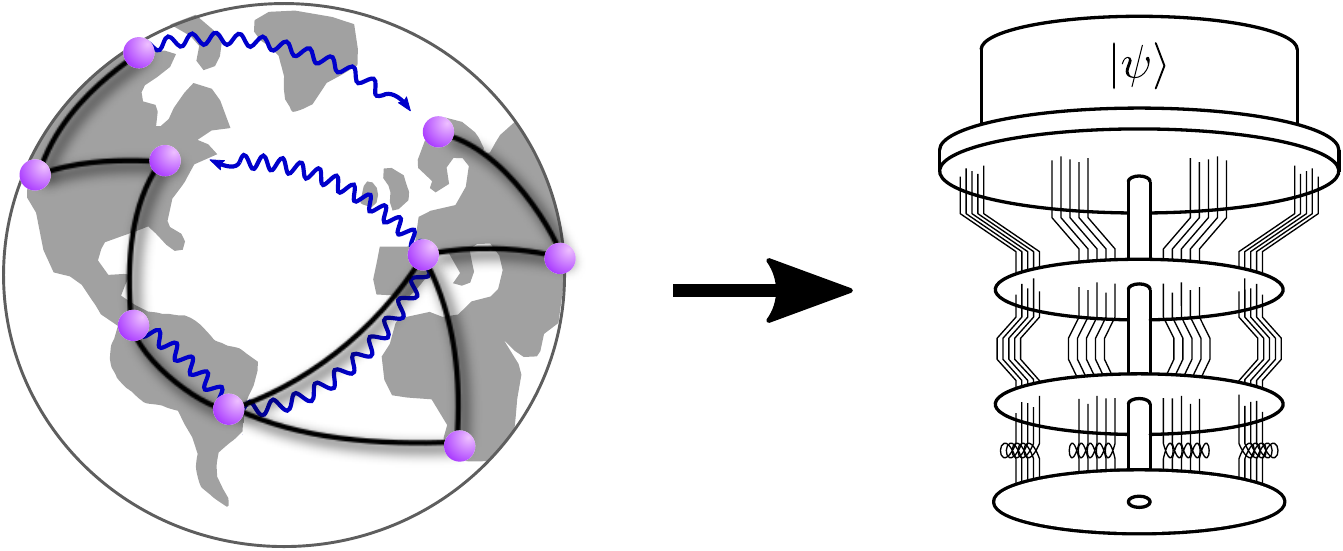}
    \caption{Illustrative general idea of the simulation of quantum communication processes using a quantum device. Any quantum network process, including noise therein, can be suitably simulated using quantum hardware, overcoming classical simulator limitations. }
    \label{fig:basicidea}
\end{figure}

In order to stress the features and benefits of the general idea presented in this work (Fig.~\ref{fig:basicidea}), we first describe our general idea of using NISQ devices for the simulation of quantum networks. We then compare our approach to other methods and argue that using quantum computers --even if they are noisy-- offers significant advantages. A more detailed analysis of the tools and strategies required for a viable quantum simulation can be found in Sec.~\ref{sec:perfect_devices}.

\subsection{General remarks on quantum simulation of quantum networks}

We start by describing the general approach we propose for the simulation of quantum networks using quantum computers. Some of these tools have been considered  \cite{nielsen2002quantum, Iten2017, qNetVO, Doolittle2023} and corresponds to quantum circuit equivalents of communication processes. The goal here is rather to point out the elements that deserve further analysis, and to propose solutions to make use of quantum hardware for simulating quantum communication protocols reliably.

A quantum network consists of multiple spatially separated nodes, where each node holds quantum devices of different kinds. Network nodes are connected to each other via classical channels (to transmit classical information), and by quantum channels to exchange quantum information. The geometry of the classical and quantum channels determines the network topology and is in principle distinct and independent from each other. Operations on devices, processes, and the transmission of quantum information are in general imperfect and noisy, and also the transmission of classical information or classical side-processing can lead to delays and errors that need to be considered. 

In a real quantum network, a protocol involves operations of devices (e.g., measurements, unitary operations, generation of photons, etc.), and transmission of quantum and classical information between them. In the simulation of such protocol, all devices are represented on a quantum computer, see Table \ref{tab:qnetwork_sim}. Devices are typically associated with quantum states of one or several qubits, and processes with operations and measurements performed on these quantum states in a quantum circuit fashion. The effect of imperfect operations, transmissions, and waiting times is simulated by performing the respective operations on the total quantum state stored in the quantum computer. We assume that the quantum computer is capable of performing standard operations, i.e., unitary operations from some universal gate set (typically arbitrary single-qubit operations and some two-qubit operation, e.g., a CNOT gate) and projective single qubit measurements \footnote{We remark that for some quantum computer hardware, also additional operations might be available that can in principle also be utilized to simplify simulations or make them more efficient. For instance, certain multi-qubit gates, POVMs, or specific non-unitary operations such as decay might be accessible. We consider later different models to describe imperfect devices, and how they can be utilized to perform simulations.}. The simulation can be done in two different ways. The first possibility is to directly simulate the time evolution of the network devices as it would occur in the real network, at all times. The second, more economic possibility, is to perform an event-driven simulation. That is, the quantum state is only updated if a relevant event occurs, by performing the associated completely positive map that describes the effective evolution after this finite time. For instance, consider an entangled state shared between two nodes in the network that is stored for some time in a quantum memory, and further processed at some later time when a classical signal arrives. One computes the effective completely positive map (CPM) that describes the evolution of the state after this time, and then only implements this CPM, without computing intermediate states, or actually waiting for this time.

{\em Quantum devices: } 
The simulation consists in describing each device (source, memory, interface, etc.) as a quantum state on the quantum computer. Typically, multiple qubits are required to provide an accurate description or model of the quantum device. In some cases, additional ancillary systems might be required to simulate specific features, or allow for an accurate, direct simulation of the underlying noise process. 

Multiple quantum devices are described by a tensor product of the corresponding quantum states, and the required number of qubits adds up. These states will become entangled when devices interact or exchange quantum information, but the required number of qubits remains the same. 

Notice that the number of required qubits for the simulation may change dynamically. This can be due to adding or removing devices or nodes from the network, or the generation of additional qubits in the protocol (e.g., a photon that is subsequently transmitted). Furthermore, measurement of qubits that then no longer participate in the process reduces the size of the quantum state to be stored and processed. For processes that only involve certain nodes or devices (which is the typical case), the total size of the required quantum state is only given by the devices that participate in the process. One may also simulate certain parts of a process using a quantum computer, which then requires fewer qubits. If different processes are separated from each other, i.e., do not involve the same qubits and devices, one maintains a tensor product structure between them. This allows one to perform simulations separately or even sequentially, thereby reducing the required size of the simulating device.

Importantly, all our methods are also applicable for $d$-dimensional systems (qudits) simulations and qudit quantum computations.

{\em Operations on devices: }
When a communication protocol is executed, operations on the devices and quantum states are performed. Typically, this will be unitary operations to manipulate quantum states. However, since in real-world applications all these operations are imperfect and noisy, this has to be accounted in the simulation. In an event-based simulation, one applies the corresponding CPM that describes the noisy operation. To do so, one needs an accurate model based on the physics of the underlying process, from which the CPM can be determined. In the simulation, one needs then a way to perform the CPM on the stored quantum state. As we describe in more detail later, this involves sequences of gates and measurements performed on the quantum computer, and typically involves some ancillary systems.     

{\em Measurements: }
Quantum communication protocols often also involve measurements, e.g., when performing entanglement purification, or when manipulating a multipartite entangled state to generate some other state on a subsystem. Similar to operations, these measurements are noisy and imperfect in reality, and can be described by a positive operator-valued measure (POVM). Again, the POVM needs to be implemented in the simulation using sequences of gates and (projective) measurements performed on an enlarged system.

{\em Quantum channels and transmission of quantum information: }
In a quantum network, quantum information is usually transmitted from one node or device to another. This is done by sending information carriers, in most cases photons, but also the transmission of other systems such as atoms is possible. Transmitted photons suffer from loss, and due to interactions with environment, the transmission will be in general imperfect. Formally, this can be described by a CPM, possibly acting on a higher dimensional system to include decay or loss. In quantum communication, these CPMs are termed noise channels, and several models have been considered and analyzed. The ideal operation is the identity in this case, and the effect of noise and imperfections is again described by the CPM. 

Similarly, one can also consider interactions between different nodes, devices or modules, which are described in a similar way as operations on devices. That is, by a CPM that acts on multiple qubits or systems, and needs to be implemented on the stored quantum state in the quantum computer.  

\begin{table}[h!]
\centering
\footnotesize
\newcolumntype{L}[1]{>{\hspace{#1}}l} 
\begin{tabular}{@{}lL{2em}@{}}
\toprule
\textbf{Network elements} & \textbf{Quantum computer simulation} \\
\midrule
Quantum node & Subset of computer registers \\
Quantum channel & CPM circuit (unitary $+$ noise) \\
Entanglement link & Entangling gate \\
Local operation & CPM circuit on node qubits \\
Measurement & POVM circuit \\
Classical communication & Conditional operations \\
Quantum memory & CPM (avoid idling times) \\
Noise and orchestration & CPM $+$ POVM circuits \\
\bottomrule
\end{tabular}
\caption{Mapping of quantum network elements to their simulation on hardware.}
\label{tab:qnetwork_sim}
\end{table}

{\em Orchestration: }
The orchestration of the processes --and also of the simulation thereof-- is typically done via a classical control plane, which steers classical and quantum devices, and classical and quantum channels. Events trigger actions, which can, e.g., be operations, measurements or transmission of classical information. The orchestration and execution of a protocol is done by exchanging classical messages, which trigger actions, including the transmission of quantum information or operations and measurements performed on quantum systems. In principle, the orchestration can also be performed via quantum states, which encode processes and actions, thereby steering the processes in a quantum way. In the simulation, this orchestration needs to be translated into an appropriate execution of CPMs and POVMs on the stored quantum state that describes network devices.

Notice that it is also possible to simulate only specific processes or parts of the quantum network using a quantum computer. In this case, the results of the simulation are integrated into a classical simulation of the quantum network. In this way, resource-consuming parts of simulation can be performed on a quantum computer, while other parts can be simulated classically.

\subsection{Comparison to previous work}

\subsubsection{Quantum simulation of noise processes}

The idea of using well-controlled quantum systems to simulate other quantum systems has been applied to simulate and investigate ground states and dynamics of physical systems and models, most notably strongly correlated systems \cite{Corboz2010, Rossmannek2023}, condensed matter systems \cite{Hofstetter2018}, field theories \cite{Freedman2002, Liu2020} or quantum chemistry \cite{McArdle2020, Cao2019}. We, in contrast, simulate interactions and protocols between complex devices and nodes in the network that themselves consist of multiple qubits. In addition, we are not interested in the dynamics or simulating a specific interaction Hamiltonian, but our simulations also include higher layers in the network stack and may include abstractions. Also, we can thus restrict ourselves to simulating completely positive maps (CPMs) rather than the full dynamics and do not need fast control or multiple intermediate operations to achieve the simulation. On the other hand, at higher layers in the network stack, protocols between complex devices involving multiple qubits and parties need to be performed, where not only gates but also sequences of measurements, classical communication, and adaptive operations, as well as operations between devices that are themselves complex quantum systems are required - which are not part in the direct simulation methods and techniques used for Hamiltonian simulation.     


Some works have analyzed the possibilities of quantum simulation using quantum devices in the context of open system dynamics, both in an analog \cite{Mostame2012, Haeffner2018, Franco2022, Daley2022, Bhowmick2023} and digital \cite{Xinlan2001, Barreiro2011, Franco14, Hu2020, Mazziotti2021, Miessen2022, Minninch2022, Wang2023, Plenio2023} fashion. While the former approach is based on the emulation of quantum systems with other quantum systems such that the simulated system evolution is governed by a desired artificial Hamiltonian or Liouvillian, the latter one makes use of other quantum systems to encode the desired information, and the simulation is carried out by appropriate quantum gates that invoke operations in a quantum computational way. Digital simulation is then closer to our work, where however, total or partial noise reduction and mitigation is used, and it requires dynamic control and high sampling costs.

Simulation of quantum channels using quantum circuits has also received a lot of attention \cite{Beny2011, Gutierrez2013, Iten2017, ChapeauBlondeau2022, Feldman2022, Mills2023, WangKai2023}. This task, of core importance for our purposes, has been mainly analyzed assuming noiseless quantum devices (or effectively achieving them via noise reduction or mitigation) \cite{Zanetti2013, Xin2017}, and also tested in real devices \cite{HeLu2017, HuLing2018, Wei2018}. We however do not require such (generally costly) mitigation approaches and, instead of trying to get rid of the inherent computer noise, we aim to take advantage of it, either directly or by transforming it. We utilize the intrinsic noise processes and reshape them in such a way that the final map is as close as possible to the desired target map. 

More broadly, the reproduction of quantum processes on quantum hardware has been considered in different contexts, fundamentally reducible to constructing quantum circuit analogues of operations—something already discussed in references \cite{nielsen2002quantum, Iten2017}.  In recent works \cite{nielsen2002quantum, Iten2017, Doolittle2023, Chen2023, qNetVO, Doolittle2024, Philip2024}, quantum simulation has been combined with variational optimization techniques (VQO) to find optimal solutions to specific network tasks or problems, where they also provide accessible and well-developed tools for simulating and variationally optimizing quantum network problems on classical or quantum computers \cite{qNetVO}. These approaches also highlight the potential of quantum hardware to aid in quantum network development by leveraging adaptive learning strategies.

The key difference with our work lies in the intended purpose of such simulations. Our approach takes a complementary perspective: we investigate how quantum computers, as noisy simulators, can be optimized or tailored to more faithfully simulate a fixed, known quantum network process. The focus is thus reversed—we do not aim to use quantum hardware to solve network problems (via VQO), but rather adapt the simulation to the device in order to study network behavior with improved accuracy compared to classical simulators or actual experiments. While conceptually simple, this perspective and its associated techniques (in particular, noise-aware tailoring) have not, to the best of our knowledge, been formally proposed or explored in prior literature.

We consider an event-based simulation of quantum networks, where we are interested in the effective description at certain times when an event occurs and we have to process the system further. This allows us to use a simplified description in terms of CPMs, and also simplified simulation processes to achieve them. Furthermore, we do not aim for correcting or mitigating noise. Our approach typically involves \textit{increasing} the amount of noise in gates and processes, since quantum channels that connect quantum devices are typically noisier than local systems, and local nodes in a network are either themselves small-scale quantum computers, or systems with increased functionality (e.g., also include an interface to photonic systems to allow for communication) and hence have a lower quality than a specialized quantum computing device. We directly utilize the existing noise and reshape its form to match one of the desired target noise processes. This is in fact a simpler and remarkably less demanding process than noise mitigation or trusting in protocols that are based on noiseless devices, and can be more easily performed using existing NISQ devices. Still, the accuracy can profit from advances in the control and quality of quantum computers.    




\subsubsection{Comparison to classical network simulation}
Quantum network classical simulators \cite{Coopmans_2021, Wu_2021, Satoh_2022, Julius2022, Chen_2023, Bassoli2020, Bel2024} have been extensively developed and widely used in the quantum networking community. These frameworks provide highly sophisticated and flexible tools to model quantum states, noisy channels, operations, and the dynamics of complex protocols including quantum repeaters \cite{Behera2019, Das2021, Ghasemi_2019}, quantum switches \cite{Dai2021}, and inter-network protocols \cite{Behera2019, Wenbo2022, Thotakura2023, Bahrani2023}. Due to their maturity and efficiency, these classical simulators have become indispensable tools for designing, benchmarking, and optimizing quantum network protocols in the current era.

However, classical simulators face fundamental computational challenges. First, the exponential growth of the Hilbert space dimension with the number of qubits or network nodes, which limits scalability for large or highly entangled quantum networks. Moreover, the simulation of noise processes that are non-Clifford or non-Pauli in nature presents additional complexity, often forcing simulators to rely on simplifying approximations or specialized techniques to remain tractable. These limitations reduce the applicability and accuracy of noise modeling in many practical scenarios.

Our work explores using near-term quantum processors (NISQ devices) as simulators for specific, known quantum network processes. While current NISQ devices have practical challenges, we study ways to improve and adapt quantum circuits to work better on these devices and get more accurate simulations. The main advantage of quantum computers is that they can naturally handle the increasing size of the quantum state space (the Hilbert space), which grows exponentially and limits classical simulators. Similarly, simulating more general noise processes that are non-Clifford or non-Pauli becomes easier and more accessible on quantum hardware.

We recognize that, for now, classical simulators are still the main practical tools for simulating quantum networks, and our approach using quantum hardware is not yet ready to replace them. Still, by showing small-scale examples and focusing on designing circuits that take noise into account, we prepare the way for future uses where quantum hardware could eventually do better than classical simulators in both speed and accuracy for some types of quantum network simulations.

\textit{System size.-- } In a quantum processor, one can treat large quantum systems, and hence, networks with multiple nodes, which is difficult or not possible with classical simulation methods. One is thereby mainly restricted by the available size of the quantum computer, where only a few ancilla qubits are required to simulate the action of different noise channels. This includes the treatment of multiplexing schemes \cite{Munro2010, MiguelRamiro2023} with shared resources (e.g., memories), the distribution of large multipartite entangled states in a network, \cite{Navascues2020,Bugalho2023} or protocols that operate on multiple copies, e.g., entanglement purification \cite{Dr2007,Riera2021} or state verification schemes \cite{Miguel2022, Riera2023}. Also for entanglement-based quantum networks, \cite{Pirker2019,MiguelRamiro2021} where large resource states are locally manipulated to obtain desired target configurations to fulfill requests, the approach is ideally suited.

\textit{Accurate error models.-- } Using quantum computers, we can include realistic models of noise in simulations. Classical simulators often have to rely on simplified noise models to manage larger systems. For example, the classical simulator QuISP \cite{Satoh_2022} uses an "error flag" model for noise, while an exact quantum noise description involves completely positive maps that can include complex effects outside Clifford or Pauli noise, or even more subtle noise features. These detailed models are important because oversimplified noise can overlook critical effects that affect the real performance of quantum network protocols.

\textit{Example.-- } We illustrate the possible advantages of a simulation using a moderate-sized quantum computer. Consider a protocol where some multipartite entangled state, say a graph state or a Dicke state, should be distributed over a large distance between $n$ nodes \cite{WallRepaters, Wrepeater}. Noise in channels and noise in local devices is assumed to be general, and not restricted to be symmetric or only of Pauli form. Hence most existing simulation methods can not be applied. Repeater protocols require operations on at least two copies of the state to perform entanglement swapping or state merging, as well as entanglement purification to increase fidelity, whereas more efficient schemes however operate on multiple copies (say $m$) in parallel. 

For simplicity, we consider only a single $m \to 1$ entanglement purification step, i.e., the iteration operates on $m$ identical copies and outputs a single purified copy. The $m$ input states need to be first generated, and the output state stored while the processor restarts the processor to generate a second purified copy. Thus to proceed with the second iteration $m$ purified copies need to be generated, meaning $R^{(1)} = 2 (nm)$ qubits are required to simulate the process using a quantum computer. Therefore, $k$ steps of entanglement purification require $R^{(k)} = 2k (nm)$ qubits. The classical simulation method needs to operate jointly on $m$ copies of a $n$ qubit state, i.e., on density matrices of the size $2^{nm} \times 2^{nm}$. Even for relatively small systems, say a state of $n=10$ qubits and operating on $m=3$ copies, this vastly exceeds the available memory and computational power, as $2^{60} \approx 10^{18}$. Multiple steps $k$ require the additional storage of the resulting state, but this overhead is insignificant as compared to the already required resources. In turn, a quantum computer of size $M=30$ suffices to perform a single step in the purification, with a linear growth in required resources when multiple steps are considered.

\subsubsection{Comparison to network benchmarking}
An alternative to simulating quantum networks—whether on classical or quantum devices—is the direct experimental construction and benchmarking of actual quantum networks \cite{Liao2022, Helsen2023}. This approach can provide the most accurate characterization of real network behavior, however, it often involves significant time and resource demands. We discuss in the following the relative strengths of simulations compared to benchmarking. 

\textit{Feasibility studies.-- } Benchmarking real networks gives direct insight but can be costly, time-consuming, or impossible with current technology, especially for large-scale or complex networks. Simulations, on the other hand, allow for rapid feasibility studies and controlled parameter variations, including access to parameter regimes beyond current experimental reach. This enables the identification of critical device qualities, bottlenecks, and potential improvements before costly network deployment.

\textit{Flexible topology.-- } While real network benchmarking is limited by physical infrastructure, simulation offers rapid exploration of diverse network topologies and configurations without physical constraints. Additionally, simulations eliminate real-time latency and waiting effects inherent in physical networks, facilitating faster iterative studies.


\textit{Simulation waiting times.-- } The impact of simulation waiting times on quantum states when sending them through a noisy channel, or keeping them in a quantum memory for a certain time, can be described by completely positive maps that are implemented on the system in one step. One only needs to take these effects into account in the simulation software, and not wait for processes to actually occur or classical signals to arrive. It is also not necessary to perform a continuous time simulation, as only the final states are of interest - corresponding to the implementation of a single CPM. Qubits in the quantum processor used for simulation are close to each other, and communication times are much smaller than between nodes that may be separated by hundreds or thousands of kilometers. Furthermore, probabilistic processes (such as photon absorption, or protocols such as entanglement purification) that in real networks lead to significant waiting times can be directly simulated, and replaced by a singe appropriate CPM.

\textit{Network and protocol optimization.--}
One can use the simulation to optimize networks, e.g., protocols or topologies. The possibility to investigate different variants in a fast and flexible way gives the opportunity to perform optimizations and find the best or improved variants that use fewer resources w.r.t. the number of devices, necessary device quality, memory or time. This approach has, e.g., been pursued in the context of entanglement-based networks \cite{Pirker2019, MiguelRamiro2021}, where resource states with reduced memory requirements were identified for a given network functionality. This helps to reduce costs and resources. 

\textit{Measurement capabilities.--} Simulations (particularly on quantum computers) can perform global or collective operations on the entire network state, enabling efficient readout strategies impossible or highly resource-intensive in physical networks, which are restricted to local node measurements and entanglement consumption for joint operations. Having access to multiple copies also allows for more efficient and improved read-out and determining properties of the network \cite{Miguel2022, Riera2023}.

It is, however, important to note the overheads associated with the readout process in each approach, typically involving process tomography \cite{Tomography2010} or, depending on the task, less demanding tools such as fidelity estimation \cite{Pallister2018}. In the quantum simulation scenario, one gains the possibility of more directly accessing the Choi matrix of the process—see below.

\textit{Complementarity with benchmarking and classical simulators.--} It is important to emphasize that benchmarking and classical simulators are mature tools addressing many of these challenges, including accurate error modeling, parameter tuning, and protocol optimization. The benchmarking methods in \cite{Liao2022, Helsen2023} are well-founded and compatible with simulation frameworks.

Our work complements these by exploring quantum hardware-based simulation approaches that can naturally scale to larger quantum states, surpassing classical simulators exponential resource requirements. This is particularly relevant for simulating general noise models and multi-copy protocols, where classical simulators rely on approximations or become intractable. However, we do not claim that near-term NISQ devices currently outperform classical simulators or benchmarking in practicality. Instead, we view our approach as a initial step toward future regimes where quantum processors may provide unique advantages in efficiency, accuracy and scalability for quantum network simulations.

\subsubsection{Comparison to quantum error mitigation approaches}
Much effort has been focused on error or noise mitigation in quantum computing, aiming to recover the output of an ideal noiseless circuit despite hardware imperfections \cite{Temme2017,GiurgicaTiron2020,Li2017,Ravi2022,Cai2023,Wallman2016,VanDenBerg2023,Ezzell2023}.  
These methods aim to suppress or extrapolate away unwanted device noise, typically at the cost of overhead in sampling or assumptions about the noise models.

Our work takes an opposite viewpoint: we aim not to remove hardware noise but to reshape, increase and exploit it so that the simulator reproduces a target noisy channel, such as a realistic memory or communication link in a quantum network, which can be expected be stronger than the inherent computer noise.  
The success criterion is therefore the fidelity with a desired noisy map, rather than the recovery of an ideal unitary.  
In many relevant cases (e.g., Pauli-type channels) the existing tools, like randomized compiling \cite{Wallman2016}, can actually help to transform the native device noise into the desired one directly, helping homogenize the hardware noise and simplifying this transformation.

Thus our approach is complementary to standard mitigation: while mitigation tries to approach noiseless computation, we deliberately tailor imperfections to emulate the real-world noise of network components, turning a limitation of NISQ devices into a resource for quantum network simulation.

%
\section{Background}
\label{sec:perfect_devices}
We present an overview of the various methods and approaches  reviewed, extended, and proposed within this work, as depicted in Fig.~\ref{fig:dilation} and Fig.~\ref{fig:enter-label}. These strategies serve as fundamental tools essential for simulating quantum network processes on quantum hardware. While our strategies are inherently general, we analyze them using simple examples yet relevant to realistic quantum communication scenarios.

\subsection{Basic concepts} \label{Sec:Background}

We first review some basic features of quantum channels and operations, and their representations that we make use of in the following.

\subsubsection{Quantum channels, operations and representations}
\label{sec.background.channels}

Any quantum operation that transforms density operators into density operators is called a quantum channel, i.e., $\hat{\mathcal{E}}: \mathcal{L} (\mathcal{H}_A) \to \mathcal{L}(\mathcal{H}_B)$ where $\mathcal{L}(\mathcal{H})$ is the set of linear operators on $\mathcal{H}$ is a quantum channel if for all density operator $\rho$ then $\hat{\mathcal{E}}(\rho)$ is also a density operator. Note that any linear, trace-preserving and completely positive quantum map is a quantum channel.  Usually, in particular in communication settings, a quantum channel is understood in a more narrow sense as a noise map that ideally corresponds to an identity operation. In the following, we do however not distinguish between quantum channels in this narrow sense and general noisy quantum operations (that ideally correspond to some unitary operation), as both are described by a completely positive map (CPM). In particular, all methods and results we describe in the following do apply to quantum channels and general CPMs.

There exist different ways of representing a quantum channel:

\begin{itemize}
    \item In the \textit{Kraus representation} a quantum channel is characterized by the so-called Kraus operators, i.e., $\hat{\mathcal{E}} = \{K_i\}$, which fulfill
    \begin{equation}
        \sum_i K_i^\dagger K_i = \id.
    \end{equation}
    The action of a quantum channel to an arbitrary state $\rho$ is given by
    \begin{equation}
        \hat{\mathcal{E}}(\rho) = \sum_i K_i \, \rho \, K_i^\dagger.
    \end{equation}
    Note the Kraus representation is not unique, as given a unitary matrix $U$ the representation with Kraus operators $\{\tilde{K}_i = \sum_{j} \langle i |U| j\rangle K_j\}_i$ is an equivalent characterization of $\hat{\mathcal{E}}$.
\item The \textit{Stinespring representation}, which follows from the Stinespring (or dilation) theorem \cite{stinespring1955positive}, states that any quantum channel can be understood as a unitary operation $\Lambda$ acting on a larger Hilbert space. A channel $\hat{\mathcal{E}}: \mathcal{L}(\mathcal{H}_A) \to \mathcal{L}(\mathcal{H}_B)$ can be conceived as a unitary $\Lambda: \mathcal{L}(\mathcal{H}_A \otimes \mathcal{H}_{\bar{A}}) \to \mathcal{L}(\mathcal{H}_B \otimes \mathcal{H}_{\bar{B}})$ that also maps the environment ($\bar{A}$) evolution, where part of the quantum state information is potentially leaked due system-environment interactions. The action of the map on an arbitrary state can be eventually described by tracing out  $\mathcal{H}_{\bar{B}}$, i.e., 
\begin{equation}
    \hat{\mathcal{E}}(\rho) = \text{tr}_{\bar{B}} \!\left[ \Lambda \left( \proj{0}_{\bar{A}} \otimes \rho \right) \Lambda^\dagger \right].
\end{equation}
The channel in the Stinespring representation is given by $\hat{\mathcal{E}} = (\Lambda, \proj{0}_{\bar{A}})$. In Appendix \ref{appendix.dilatation.1} we show how the Stinespring and Kraus representations uniquely relate.

\item The \textit{Choi representation} of a quantum channel is defined by the Choi state of the channel, denoted as $\Phi_{\hat{\mathcal{E}}}$. It is expressed as:
\begin{equation}
    \Phi_{\hat{\mathcal{E}}} = \id \otimes  \hat{\mathcal{E}}  \left( \proj{\Phi} \right),
\end{equation}
where $\ket{\Phi}$ represents a bipartite maximally entangled state given by $\ket{\Phi} = \sum_{i=0}^{d-1} \ket{i, i} / \sqrt{d}$. The Choi state offers a comprehensive description of the channel, enabling direct extraction of the Kraus operators. As $\Phi_{\hat{\mathcal{E}}}$ is a density operator, it can always be decomposed as:
\begin{equation}
    \label{eq:ensemble.decomposition}
    \Phi_{\hat{\mathcal{E}}} = \sum_{i=0}^{d^2-1} p_i \proj{\psi_i},
\end{equation}
where ${ p_i }$ forms a probability distribution and $\{ \ket{\psi_i} \}$ represents a set of normalized states. By expressing $\ket{\psi_i} = \id  \otimes  \Omega_i \ket{\Phi}$, where $\bra{k} \Omega_i \ket{l} = \bracket{k,l}{\psi_i}$, it is evident that ${ K_i = \sqrt{p_i} , \Omega_i }$ constitutes a Kraus representation of $\hat{\mathcal{E}}$. Notably, this decomposition aligns with the non-uniqueness of the Kraus representation, as observed in Eq.~\eqref{eq:ensemble.decomposition}. We denote the Kraus rank of $\hat{\mathcal{E}}$ as $r$, which is determined by the rank of $\Phi_{\hat{\mathcal{E}}}$, corresponding to the minimum number of elements in a Kraus representation of the channel $\hat{\mathcal{E}}$.

\item The \textit{Liouville Superoperator representation} of a quantum channel encapsulates the complete description of the channel in a matrix known as the Liouville superoperator, defined as (note we describe these channel matrices with no-hat notation):
\begin{equation}
\mathcal{E} = \sum_i K_i^* \otimes K_i.
\end{equation}
This representation is particularly significant as it facilitates channel concatenation through matrix multiplication. Specifically, the Liouville Superoperator of the concatenation of multiple channels corresponds to the matrix product of their respective Liouville Superoperators, i.e., if $\hat{\mathcal{E}}_3 = \hat{\mathcal{E}}_2 \circ \hat{\mathcal{E}}_1$, then $\mathcal{E}_3 = \mathcal{E}_2 \cdot \mathcal{E}_1$. Moreover, the action of a channel on an arbitrary state $\rho$ is given by $\mathcal{E} \ket{\rho} \! \rangle$, where $\ket{\rho} \! \rangle = \sum_{i,j} \bra{i} \rho \ket{j} \ket{ij}$ represents the vectorized form of $\rho$.

For a given quantum channel, its Choi matrix and its Liouville superoperator are connected through a reshuffling of their matrix elements, enabling easy conversion between the two representations.
\end{itemize}

The distance between two channels, $\hat{\mathcal{E}}_1$ and $\hat{\mathcal{E}}_2$, can be evaluated using various metrics. In this work, we utilize the Choi fidelity, i.e.,
\begin{equation}
    F(\mathcal{E}_1,\mathcal{E}_2) = \left( \text{tr} \sqrt{\sqrt{\Phi_{\mathcal{E}_1}} \Phi_{\mathcal{E}_2} \sqrt{\Phi_{\mathcal{E}_1}}} \; \right)^2.
    \label{eq:channel_fid}
\end{equation}
The channel infidelity between two channels, $\hat{\mathcal{E}}_1$ and $\hat{\mathcal{E}}_2$, is defined as $1-F(\mathcal{E}_1,\mathcal{E}_2)$. 

Observe that, although experimentally unfeasible, this quantity can actually be accessed on a quantum computer, where an entangled state using an ancilla can be initially prepared, letting the ancilla untouched until the end. The noise introduced by the additional entangling gates and the additional ancilla required can be incorporated in the simulation to directly access the channel fidelity or the process matrix in a more efficient way, thereby reducing the overheads typically required, e.g., for full process tomography. Alternatively, process fidelity estimation techniques could be used in practice to test the quality of the simulations.

\subsection{Quantum hardware noise } \label{sec:back_model}

Although the strategies we introduce in this work can be conceived as tools for completely general quantum network simulation, in the results we present, we make specific assumptions about the noise models of the quantum devices. Although these assumptions are inspired by realistic sources of noise observed in current quantum hardware \cite{Resch2021, Georgopoulos2021}, our methods are not restricted to them and are just chosen for illustrative purposes.

\subsubsection{Noise models}

\begin{figure}
    \centering
    \subfloat[Gate noise model]{\includegraphics[width=\columnwidth]{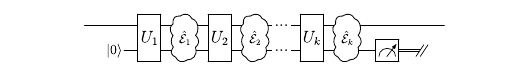} \label{fig:noise_modelsa}} \hfil \\
    \subfloat[Block noise model]{\includegraphics[width=\columnwidth]{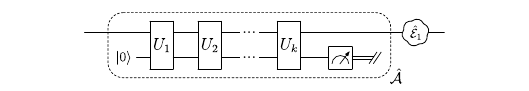} \label{fig:noise_modelsb}} \hfil 
    \caption{  Basic schemes considered to model imperfect hardware operations for the simulations. (a) Each elementary quantum gate is modeled as the perfect gate application followed by some noisy channel. For instance, considering the amplitude damping as reference circuit $\hat{\mathcal{A}}$, one finds $U_1= R_y(\theta) = \exp\{-\text{i} \frac{\theta}{2} Y \}$, $U_2=U_k=\text{CNOT}_{21}$. (b) The noise is modeled as the perfect circuit implementation --corresponding to a quantum map $\hat{\mathcal{A}}$--, followed by some noisy channel. }
    \label{fig:noise_models}
\end{figure}

We consider two distinct methodologies for modeling hardware noise within quantum computers during simulations:

\underline{Gate noise model}:  The first model we consider, commonly used in the literature, involves analyzing noisy operations in the simulation by representing the noise of each elementary gate in the quantum computer as depicted in Fig.~\ref{fig:noise_modelsa}. Specifically, we model each noisy gate by first implementing the ideal version of the gate, followed by the introduction of noise channels affecting each register involved. Typical noise channels considered include dephasing, depolarizing, or amplitude damping, which are prevalent in current quantum devices \cite{Resch2021, Georgopoulos2021}.

\underline{Block noise model}:  Alternatively, we also consider a compacted model that represents the noise of a quantum circuit --corresponding to a particular quantum map-- by employing the (whole quantum circuit) ideal simulation of the map followed by a noise channel, structured in a block format, as illustrated in Fig.~\ref{fig:noise_modelsb}. Practically, this model can prove highly beneficial when direct knowledge of the quantum gates' inherent noise is unavailable. In such cases, one can model the noise effect as the ideal map implementation followed by an encompassing noise channel, effectively accounting for all noise sources during computation.

\subsubsection{Noisy channels }
Regarding the specific noise sources under consideration, as previously noted, we primarily examine three types of noise:

\textit{Dephasing noise: }This type of noise is a significant factor contributing to the loss of coherence in quantum systems. It manifests in various quantum scenarios, such as imperfections in optical fibres or fluctuations in electromagnetic fields. Mathematically, it is defined as follows: 
\begin{equation}
    \hat{\mathcal{E}}(\rho)= p \rho + (1-p) \sigma_z \, \rho  \, \sigma_z, 
\end{equation}
where $\sigma_z$ denotes the $Z$-Pauli operator.

\textit{Depolarizing noise.-- }  Also referred to as white noise, depolarizing noise holds particular relevance as it often enables the analysis or estimation of worst-case scenarios. Its expression can be represented as: 
\begin{equation}
    \hat{\mathcal{E}}(\rho) = p \rho + \frac{1-p}{3} (\sigma_z \rho  \,\sigma_{z} + \sigma_x \rho \, \sigma_x + \sigma_y \rho \, \sigma_y).
\end{equation}
Alternatively, it can be equivalently expressed as: $\hat{\mathcal{E}}(\rho) = p' \rho + (1-p') \rho \id/2$,
where $p'=(4p-1)/3$. This transformation maps the state to a completely mixed state with probability $(1-p')$, implying the loss of all state information with that probability. Notably, the depolarizing channel represents a comprehensive scenario for single-qubit channels, as any other channel or state can be transformed into a depolarizing form through random operations, a process known as depolarization.

\textit{Amplitude damping noise.-- } This noise type is a primary example of a non-unital noisy channel, providing a representation of important phenomena such as spontaneous emission. The Kraus operations defining this noise are as follows:

\begin{equation}
\begin{aligned}
    \label{eq:dampingKrauses}
    K_0 & = \proj{0} + \sqrt{1-\gamma} \proj{1}, \\
    K_1 & = \sqrt{\gamma} \ketbra{0}{1}.
\end{aligned}
\end{equation}

We remark that while we employ these assumptions for analyzing examples and demonstrating specific results, our methods remain entirely general. Indeed, they can be seamlessly integrated into a broader variational framework (refer to Sec.~\ref{sec:Variational}) to ensure independence from specific noise forms and models.

\section{Tools for the simulation of channels, noisy operations and measurements} \label{sec:approaches}

A crucial ingredient for achieving a reliable simulation of quantum network processes lies in the ability to simulate quantum channels, operations, and measurements with high fidelity. The methodologies discussed here encompass the necessary tools for simulating noise in a quantum computer across various models, settings, and assumptions.

In this section, we elaborate on the assumption of a perfect quantum device, where unitary quantum gates, projective measurements, and state preparation processes remain unaffected by any form of noise. Despite the idealistic nature of this assumption, it serves as a foundational framework for understanding the implementation and manipulation of quantum channels, noisy operations, and measurements --whether described by completely positive maps (CPMs) or positive operator-valued measures (POVMs)-- within a quantum circuit architecture. Strategies for managing and leveraging inherent computational noise are addressed in Sec.~\ref{sec:noisy_devices}.

We delve into the analysis of how an arbitrary $n$-qubit quantum channel or operation can be implemented (simulated) within a perfect quantum device. We presume a quantum computing system capable of executing only unitary gates, projective measurements, and stochastic mixtures of both. Any channel becomes implementable if the quantum processor manipulates a quantum system possessing a sufficiently large Hilbert space (number of qubits). Within this section, we examine the spatial overhead $R_n$ necessary to implement a general $n$-qubit channel, defined as the number of ancillary qubits required by the quantum processor, i.e., $R_n = \log_2 D - n$, where $D$ represents the dimension of the Hilbert space of the quantum computer. Given that the state of an $n$-qubit system is equivalent to that of a qudit system with dimension $d=2^n$, we restrict our focus in this section to single-qudit channels and operations.

We explore two distinct approaches: firstly, an ancillary-assisted quantum computer configuration, where the quantum processor comprises $N$ qubits, thereby setting $D = 2^N$; and secondly, a scenario where the processor can make use of an arbitrarily large qudit system, denoted as $D \in \mathbbm{N}$.

\begin{figure}
    \centering
    \subfloat[Approach I. Ancilla-assisted]{\includegraphics[width=\columnwidth]{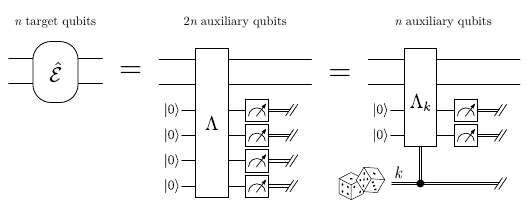}\label{fig:dilationa}} \hfil
    \subfloat[Approach II. Extended qudit]{\includegraphics[width=\columnwidth]{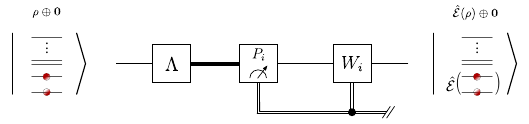}\label{fig:dilationb}}
    \caption{\label{fig:dilation} Tools for channel simulation in a perfect quantum device (a) \textit{Approach I. Ancilla-assisted.} Any $n$-qubit channel $\hat{\mathcal{E}}$ can be implemented as a unitary matrix $\Lambda$ acting on a $3n$-qubit system. If stochastic mixtures of unitaries can be implemented, only a $2n$-qubit ancillary system is sufficient. (b) \textit{Approach II. Extended qudit.} Encoding a qudit state in a super-qudit, a channel  can also be implemented through unitary gates and a projective measurement into the super-qudit.}
\end{figure}

\subsection{Approach I. Ancilla-assisted}
\label{sec.dilation.1}

In the first approach, which we denote as the \textit{ancilla-assisted}, simulating quantum channels is achieved directly through unitary gates and projective measurements. According to the Stinespring dilation theorem (refer to Sec.~\ref{sec.background.channels}), any quantum channel or completely positive map (CPM) $\hat{\mathcal{E}}$ can be represented as a unitary transformation $\Lambda$ acting on an extended system. By simulating this unitary transformation and subsequently tracing out the additional ancillary subsystems, one can effectively implement any channel within a quantum device, as illustrated in Fig.~\ref{fig:dilationa}. Consequently, in this approach, the spatial overhead is determined by  $R_n = \log_2 \!\left( d_a \right)$, where $d_a$ denotes the dimension of the ancillary system.

In Appendix~\ref{appendix.dilatation.1}, we provide a detailed procedure for determining the unitary gate $\Lambda$ corresponding to any given quantum channel. The dimension of $\Lambda$ is given by $dim(\Lambda) = r d$, where $r$ represents the Kraus rank of the channel. As the Kraus rank of a qudit channel is upper-bounded by $d^2$, this method enables the implementation of any $n$-qubit channel with an overhead of $R_n = 2n$. Consequently, for a system comprising $n$ qubits, the quantum processor should be capable of manipulating $N = 3n$ qubits.

We initially investigate the minimum ancilla requirements for stimulating a CPM using extreme channels. The set of quantum channels is convex, meaning that any arbitrary channel $\hat{\mathcal{E}}$ can be expressed as a convex combination of extreme channels. Mathematically, this can be represented as:
\begin{equation}
    \label{eq.generalized.extreme.channel}
    \hat{\mathcal{E}}(\rho) = \sum_k p_k \, \hat{\mathcal{E}}^{\text{ext}}_k(\rho),
\end{equation}
where $0< p_k \leq 1$, $\sum_k p_k =1$. Here, $\hat{\mathcal{E}}^{\text{ext}}_k$ denotes the extreme channels, which are those that cannot be expressed as convex combinations of other quantum channels \cite{CHOI1975285, BETHRUSKAI2002159, Wang_2015}. Therefore, one can infer from Eq.~\eqref{eq.generalized.extreme.channel} that $\hat{\mathcal{E}}$ can be realized by implementing $\hat{\mathcal{E}}^{\text{ext}}_k$ with probability $p_k$. 

The Kraus rank of any extreme channel is at most $d$, where $d$ represents the dimension of the Hilbert space. Therefore an ancillary system of dimension $d$ is sufficient to implement any $\hat{\mathcal{E}}^\text{ext}$ and, thus, any quantum channel $\hat{\mathcal{E}}$. This insight leads to an overhead of $R_n = n$, meaning that the minimal size can be reduced to $N = 2n$ qubits.

Our objective is however to further minimize the resource requirements for channel simulation. In this regard, employing techniques such as stochastic maps or classical mixtures of unitary operations can facilitate the implementation or approximation of various types of quantum channels even without the need for ancillary systems. In the subsequent discussion, we show relevant cases where utilizing stochastic mixtures of unitary gates enables a reduction in the size of the required quantum processor.

\subsubsection{Relevant cases}

We elaborate on these distinctive features for various types of completely positive maps (CPMs), where the need for ancillary systems required to simulate them can be minimized or even disregarded.

\textit{Unital channels.-- } One notable category is that of mixed-unital channels, denoted as $\hat{\mathcal{U}}$, which consist of convex combinations of unitary gates. Mathematically, they are expressed as:
\begin{equation}
    \hat{\mathcal{U}}(\rho) = \sum_k p_k \, U_k \, \rho \, U_k^\dagger,
\end{equation}
where $\{ p_k \}$ form a probability distribution and $U_k$ represents a unitary gate. These channels can be implemented by applying gate $U_i$ with probability $p_i$, without the necessity for ancillary systems. Noteworthy noisy channels falling into this category include bit flip noise ($\hat{\mathcal{X}} = \{ p \, \id, (1-p) X\}$), phase damping ($\hat{\mathcal{Z}} = \left\{ \sqrt{ p_k } \, e^{i \theta_k Z } \right\}_k$), and depolarizing noise, discussed in Sec.~\ref{Sec:Background}.

\textit{Projective measurement channels.-- } Another class of channels that can be realized without ancillary systems are those that can be decomposed into a projective measurement followed by a correction operation. These channels possess a Kraus representation of the form $\hat{\mathcal{M}} = \{ U_i P_i \}$, where $P^2_i = P_i$ denotes a projector. Implementing channel $\hat{\mathcal{M}}$ involves performing a measurement described by $\{ P_i \}$  followed by the correction operation $U_i$, eventually discarding the measurement outcome. An important example of this type is the quantum reset channel ($\hat{\mathcal{R}}$), which is a specific instance of an extreme channel that can be implemented without requiring ancillary systems, although this is not generally true for all extreme channels.

\textit{Erasure channel.-- } The quantum erasure channel models scenarios where the physical particle encoding quantum data is lost with a certain probability $p$, such as photon loss in optical systems. This channel transforms a qubit state $\rho$ as follows:
\begin{equation}
    \hat{\mathcal{N}}_p (\rho) = p \, \rho + (1-p) \proj{e} \in \mathcal{L}\left(\mathbbm{C}^{d+1}\right),
\end{equation}
where $\ket{e}$ is an orthogonal state to $\rho$. The erasure channel can be simulated with a single ancillary system in the pure state $\ket{0}_a$, by leaving the state untouched with probability $p$, and projecting the system to the state $\ket{0}\ket{1}_a$ with probability $(1-p)$, i.e.,
\begin{equation}
    \hat{\mathcal{N}}_p (\rho) = p \, \rho \otimes \proj{0} + (1-p) \proj{0} \otimes \proj{1},
\end{equation}
In this scenario, the ancillary states are not traced out but must be retained throughout the computation.

An alternative method to simulate the quantum erasure or loss channel involves utilizing an additional energy level—effectively embedding the qubit into a qutrit system \cite{Gu2025}. This approach allows to simulate the erasure process more naturally.

\subsection{Approach II. Extended qudit}
\label{sec.dilation.2}

We propose and explore here the benefits derived from employing higher-dimensional systems, referred to as the \textit{extended qudit} approach, where different levels or subsystems of the qudits can play the role of ancillary registers. We investigate the resource requirements for simulating arbitrary quantum channels, noting that using this strategy with various dimensions of qudits can be particularly advantageous for simulating certain channels.

We advocate for the simulation of quantum channels utilizing \textit{qudit} systems. Although quantum information is still encoded in qubits, these qubits are embedded within a subspace of the $d$-level system, which we denote as the \textit{data subspace}. Consequently, the quantum processor can utilize the additional Hilbert space to dilate any quantum channel into a routine comprising unitary operations and projective measurements. Leveraging qudit systems can lead to more efficient channel implementations, as only single qudit operations are required instead of two-qubit operations (qubit-ancilla). Importantly, experimental research has demonstrated that individual qudits can be fully manipulated with comparable accuracy to a single qubit \cite{Ringbauer2022, Hrmo2023}. Additionally, as detailed below, employing qudit systems can reduce the simulation overheads compared to the ancilla-assisted approach.

We proceed to demonstrate how, given an arbitrary quantum channel, one can derive a routine $\hat{\mathcal{P}}$ that implements it in a qudit quantum computer, as illustrated in Fig.~\ref{fig:dilationb}. For mathematical convenience, we assume that quantum information is encoded in a qudit state $\rho \in \mathcal{L}(\mathbbm{C}^d)$. This qudit state is implemented within an extended qudit system of larger dimension, where the state of the entire system is represented by $\rho \oplus \textbf{0} \in \mathcal{L}(\mathbbm{C}^D)$.

\textit{Routine $\hat{\mathcal{P}}$.-- } Given a qudit channel $\hat{\mathcal{E}} = \{K_i\}_{i=0}^{r-1}$, we initially apply a unitary gate $\Lambda$ that encodes $\{ W^\dagger_i K_i \, \rho \, K_i^\dagger W_i \}_{i=0}^{r-1}$ in different orthogonal subspaces of the Hilbert space of the extended qudit system, where $W_i$ is a unitary gate. Subsequently, we perform a projective measurement  $\{ P_i \}_{i=0}^{r-1}$ into each of these subspaces. This measurement leaves the state of the entire system in the state $\frac{1}{p_i} W_i^\dagger K_i \rho K_i^\dagger W_i$ in the corresponding subspace with probability $p_i = \text{tr}[P_i \Lambda (\rho \oplus \bo{0})\Lambda^\dagger]$. We then implement a correction operation $\bar{W}_i$, which first returns the state of the qudit back into the data subspace and then inverts $W_i^\dagger$. At this stage, the state of the system is given by
\begin{align*}
    \hat{\mathcal{P}}_i(\rho \oplus \bo{0}) & = \frac{1}{p_i} \bar{W}_i \, P_i \, \Lambda \left( \rho \oplus \bo{0} \right) \Lambda^\dagger \, P_i \, \bar{W}^\dagger_i \\
    & = \frac{1}{p_i} \big(K_i \, \rho \, K_i^\dagger \big) \oplus \bo{0}.
\end{align*}
where $\hat{\mathcal{P}}_i$ refers to the branch of routine $\hat{\mathcal{P}}$ where outcome $P_i$ is obtained. Finally, the measurement outcome is erased, leading to a convex mixture of all branches, i.e.,
\begin{align*}
    \hat{\mathcal{P}}(\rho \oplus \bo{0}) = \sum_{i=0}^{r-1} p_i \, \hat{\mathcal{P}}_i(\rho \oplus \bo{0}) = \hat{\mathcal{E}}(\rho) \oplus \bo{0}.
\end{align*}
Refer to Appendix~\ref{appendix.dilatation.D} for detailed explanations.

The dimension of the subspace required to encode $W_i^\dagger K_i \, \rho \, K_i^\dagger W_i$ is given by $\text{rank}(K_i) = \kappa_i$. Therefore, the quantum channel can be simulated if $D \geq \sum_{i=0}^{r-1} \kappa_i$. It is noteworthy that $\sum_{i=0}^{r-1} \kappa_i \leq r d$, indicating that the utilization of qudits can result in a reduced spatial overhead compared to ancillary systems.

\textit{Example.--} Consider the amplitude damping channel $\hat{\mathcal{A}}$ from Eq.~\eqref{eq:dampingKrauses}. In this case, the sum of the ranks of the Kraus operators is 3, indicating that it suffices to implement the qubit in a 3-level system. Thus, the input state is given by $dim(\rho \oplus 0) = 3\times 3$. First, we apply the unitary gate
\begin{equation}
    \Lambda =
    \begin{pmatrix}
        1 & 0 & 0 \\
        0 & \sqrt{1-\gamma} & -\sqrt{\gamma} \\
        0 & \sqrt{\gamma} & \sqrt{1-\gamma}  \\
    \end{pmatrix},
\end{equation}
followed by the projective measurement with $P_0 = \proj{0} + \proj{1}$ and $P_1 = \proj{2}$, along with the corresponding correction operations $\bar{W}_0 = \id$ and $\bar{W}_1 = X^\dagger_3$ where $X_D^\dagger \ket{k} = \ket{(k+1) \text{ mod } D}$.  As a result, we obtain the state $\hat{\mathcal{A}}(\rho) \oplus 0$. Therefore, the overhead is given by $R_1 = 1 - \log_2 3 \approx 0.58$. 

On the contrary, if we utilize the ancilla-assisted approach, an ancilla qubit is required, resulting in an overhead of $R_1 = 1$. This comparison illustrates an example of the advantage of employing the extended qudit approach, as it leads to a reduction in overhead compared to the ancillary-assisted approach.

\subsection{POVM simulation}

The approaches outlined above for simulating noisy channels and operations can also be extended to simulate noisy measurements, as described by a Positive Operator-Valued Measure (POVM).  Given a POVM \cite{nielsen2002quantum} $\mathcal{O} =\left\{ O_i \geq 0 \right\}_{i=0}^{p-1}$, it can always be implemented as a Von Neumann measurement acting on a larger Hilbert space. If the input state is considered to be given by $\rho \oplus \bo{0}$, then $\mathcal{O}$ can be performed by implementing channel $\hat{\mathcal{E}} = \left\{ \sqrt{O_i} \right\}_{i=0}^{p-1}$ as described previously.  However, in this case, the outcome of the projective measurement on the qudit system is not erased but learned. Therefore, the dimension of the extended qudit system must satisfy  $D \geq \sum_{i=0}^p \text{rank}(O_i)$.

\section{Network simulation with noisy quantum devices} \label{sec:noisy_devices}

In any quantum computing hardware, NISQ devices in particular, noise and decoherence are inevitable factors that affect the reliability of quantum computations. When simulating quantum communication or network processes on such devices, it becomes imperative to incorporate these noisy characteristics. Achieving a faithful simulation of quantum communication processes on quantum hardware necessitates the utilization of diverse tools and techniques. These tools enable the replication of quantum schemes within a quantum circuit while effectively managing -or even leveraging- the inherent noise of the system. As already stressed before, our evaluations typically focus on simulating specific quantum channels, with a designated target denoted as $\hat{\mathcal{E}}_{\text{t}}$, representing the desired channel to be implemented on the quantum hardware. While an ideal, noiseless quantum computer could implement the channel with perfect fidelity using the techniques discussed in Sec.~\ref{sec:approaches}, imperfect quantum hardware necessitates the development of strategies to mitigate noise effects, exploit them, or transform certain quantum channels. We primarily rely on static noise models. In practice, however, quantum device noise often exhibits temporal fluctuations due to environmental factors such as temperature drift, vibrations, and electromagnetic interference. While our framework does not simulate noise dynamically at every intermediate time step, it remains applicable by focusing on reproducing the effective initial-to-final quantum process, which can be captured as a single CP map.    In this context, we propose several general methods to address these challenges and evaluate their effectiveness through relevant examples.

Firstly, we introduce the \textit{building-block} method, where circuits simulating the desired channels are conceived as fixed building blocks. In this approach, the inherent structure of the simulation circuit -such as the sequence of gates- is immutable and inaccessible for direct modification. However, the circuit remains vulnerable to noise, resulting in imperfect implementation. To counteract this, we introduce additional channels, also implemented within a quantum circuit framework and subject to noise. These additional channels can be strategically applied before, after, or via classical mixture to compensate for the noise effects and approximate the overall circuit to the desired implementation.

Secondly, we explore the \textit{tailored circuits} method, where circuits intended for channel simulation are not fixed. Here, one can adjust the circuits to approximate the desired channel as closely as possible, considering hardware imperfections. We demonstrate instances where the direct implementation of channel simulation circuits, which is optimal in noiseless scenarios, becomes suboptimal when noise is accounted for. Additionally, we present various scenarios with differing flexibility ranges for optimizing the quantum circuit.

Lastly, we propose an extension that can complement or integrate with the aforementioned strategies (and tools from Sec.~\ref{sec:approaches}), leveraging variational quantum optimization techniques. These techniques enable the simulation of a quantum channel with only accessible input parameters, without requiring detailed knowledge of the simulation's internal workings. By maximizing the fidelity of the simulated process through appropriate optimization of these input parameters, we achieve an approximation of the desired simulation directly. We refer to this strategy as the \textit{Variational black-box} method. It is essential to address preparation and measurement errors and design efficient procedures for measuring channel fidelity.

Real devices can be seamlessly integrated into simulations either via direct interface replacement or through benchmarking to obtain an optimized description within a specified error model.

\begin{figure}
    \centering
    \subfloat[Method 1. Building-block channels]{\includegraphics[width=\columnwidth]{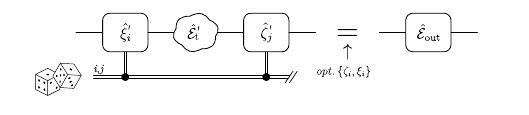}\label{fig:enter-label_a}} \hfil
    \subfloat[Method 2. Tailored circuit channels]{\includegraphics[width=\columnwidth]{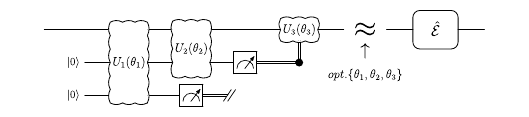}\label{fig:enter-label_b}} \hfil
    \subfloat[Method 3. Variational black-box optimization]{\includegraphics[width=\columnwidth]{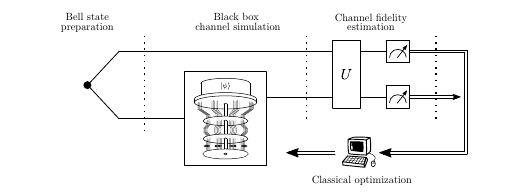}\label{fig:enter-label_c}} \hfil
    \caption{\label{fig:enter-label} Strategies for transforming and harnessing inherent NISQ devices noise. (a) \textit{Method~1. Building-block channels.} In this approach, the circuits simulating channels are inaccessible, and noise is mitigated by constructing and applying additional channels before or after.   (b) \textit{Method~2. Tailored circuit channels.}  This method involves tuning the circuits—including gates and parameters therein—used to implement channels. This tuning allows for the transformation, compensation, or leveraging of computer noise. (c) \textit{Method~3. Variational black-box optimization.}  In this technique, the quantum device executes a circuit, where some input parameters are optimized variationally to match or approximate the desired simulation. Input parameters are iteratively updated via classical optimization based on a figure of merit without prior knowledge of the underlying noise or circuit structure. }
\end{figure}

\subsection{Method 1. Building-block channels} \label{sec: Method3b}

The first strategy we propose is denoted as the building-block method, illustrated in Fig.~\ref{fig:enter-label_a}. In this method, the circuits used to simulate quantum channels are treated as fixed building blocks. These circuits, representing the fundamental elements of the simulation process, are not directly modifiable, including the sequences of gates and operations. However, it is possible to introduce additional channels within the quantum circuit framework. These supplementary channels, also susceptible to noise, are strategically applied in various configurations. For instance, they can be applied before an input channel, after it, or both (referred to as \textit{interleaved}), with the aim of achieving a satisfactory approximation of the desired target channel ${\hat{\mathcal{E}}_\text{t}}$. Note the target channel equals the noiseless implementation of the input channel. Additionally, the method allows for the consideration of non-trivial classical mixtures of these channels applied with different probabilities. It is crucial to note that directly inverting a quantum channel is not possible. 

Given an input and a target quantum channels (identical in a noiseless hardware scenario), the building-block method implements 
\begin{equation}
     \hat{\mathcal{E}}_{\text{out}} = \sum_{i,j} p_{ij} \, \hat{\zeta}'_i \circ \hat{\mathcal{E}}'_{\text{t}} \circ \hat{\xi}'_j,
\end{equation}
such that maximizes the channel fidelity between the output and target channels $F(\hat{\mathcal{E}}_{\text{out}},\hat{\mathcal{E}}_t)$, such that $\hat{\mathcal{E}}_{\text{out}} \approx \hat{\mathcal{E}}_{\text{t}}$, where $\{p_{ij}\}$ defines a probability distribution and $\{ \hat{\zeta}_i, \hat{\xi}_j \}$ are quantum channels. The prime notation, i.e. $\hat{\xi}'$ indicates the channel implementation is subjected to computer noise that deviates from the ideal channel implementation $\hat{\xi}$. This deviation can be considered by the method when optimizing over the construction of the channels $\{ \hat{\zeta}_i, \hat{\xi}_j \}$.  In the examples shown below we model the hardware noise using the models introduced in Sec. \ref{sec:back_model}.

Utilizing the superoperator representation, as discussed in Sec.~\ref{sec.background.channels}, channel concatenation can be expressed as matrix multiplication.  With this formalism, the problem reduces to finding matrices ${\zeta_i, \xi_i}$ that maximize channel fidelity while considering noise (i.e., accounting for their imperfect implementation ${{\zeta}'_i, {\xi}'_i}$). These matrices must represent valid physical quantum channels, meaning that the corresponding channels (or Choi matrices, see Sec.~\ref{sec.background.channels}) ${\hat{\zeta}_i, \hat{\xi}_i}$ adhere to positivity, trace preservation, and complete positivity properties. Due to the analytical complexity of the problem, we rely on numerical analyses based on various optimization techniques.  We emphasize that our examples focus on specific scenarios, but the strategies are broadly applicable.

\begin{figure}
    \centering
    \subfloat[]{\includegraphics[width=0.9\columnwidth]{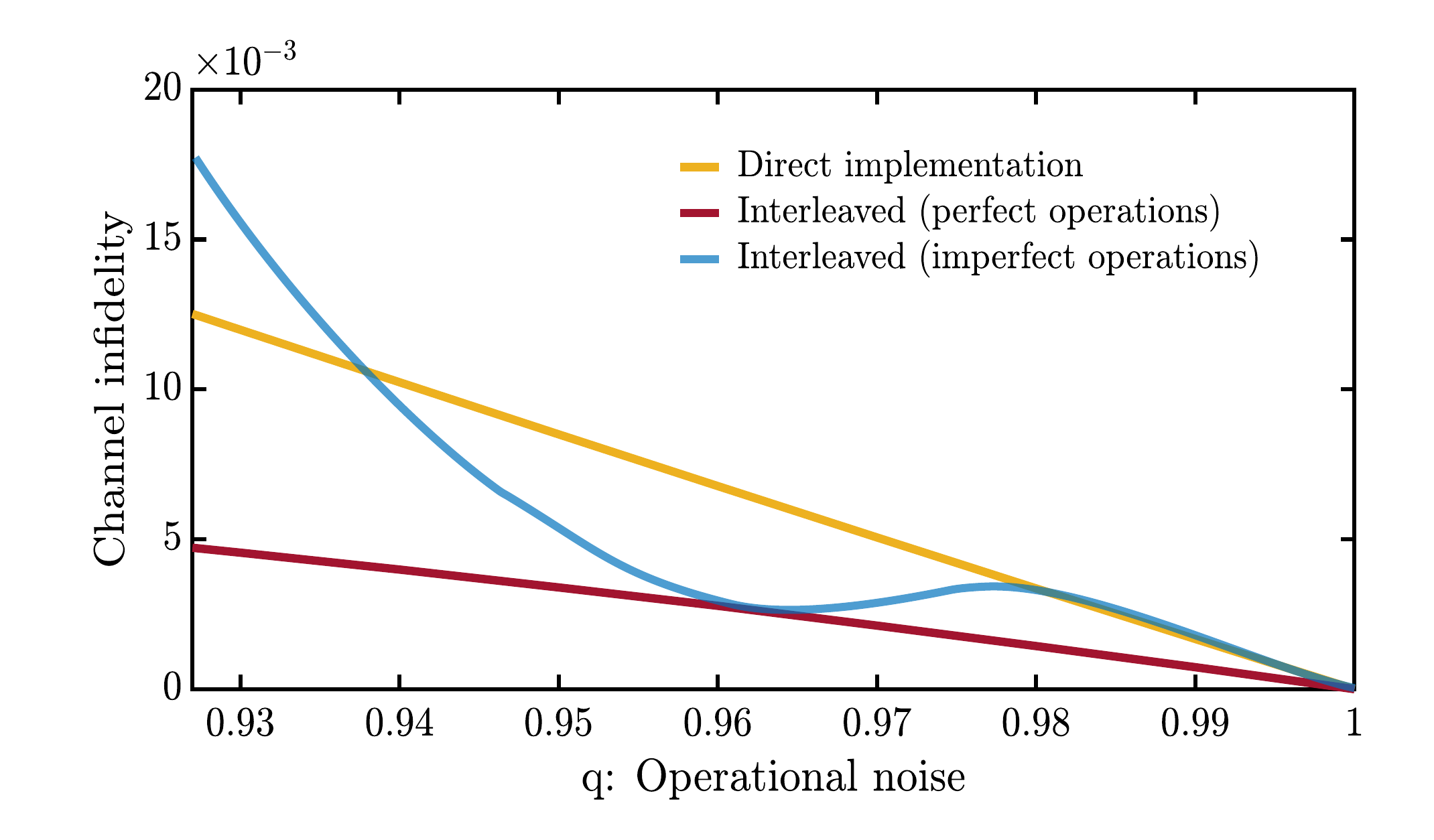} \label{fig:depo_blocks}} \\
    \subfloat[]{\includegraphics[width=0.9\columnwidth]{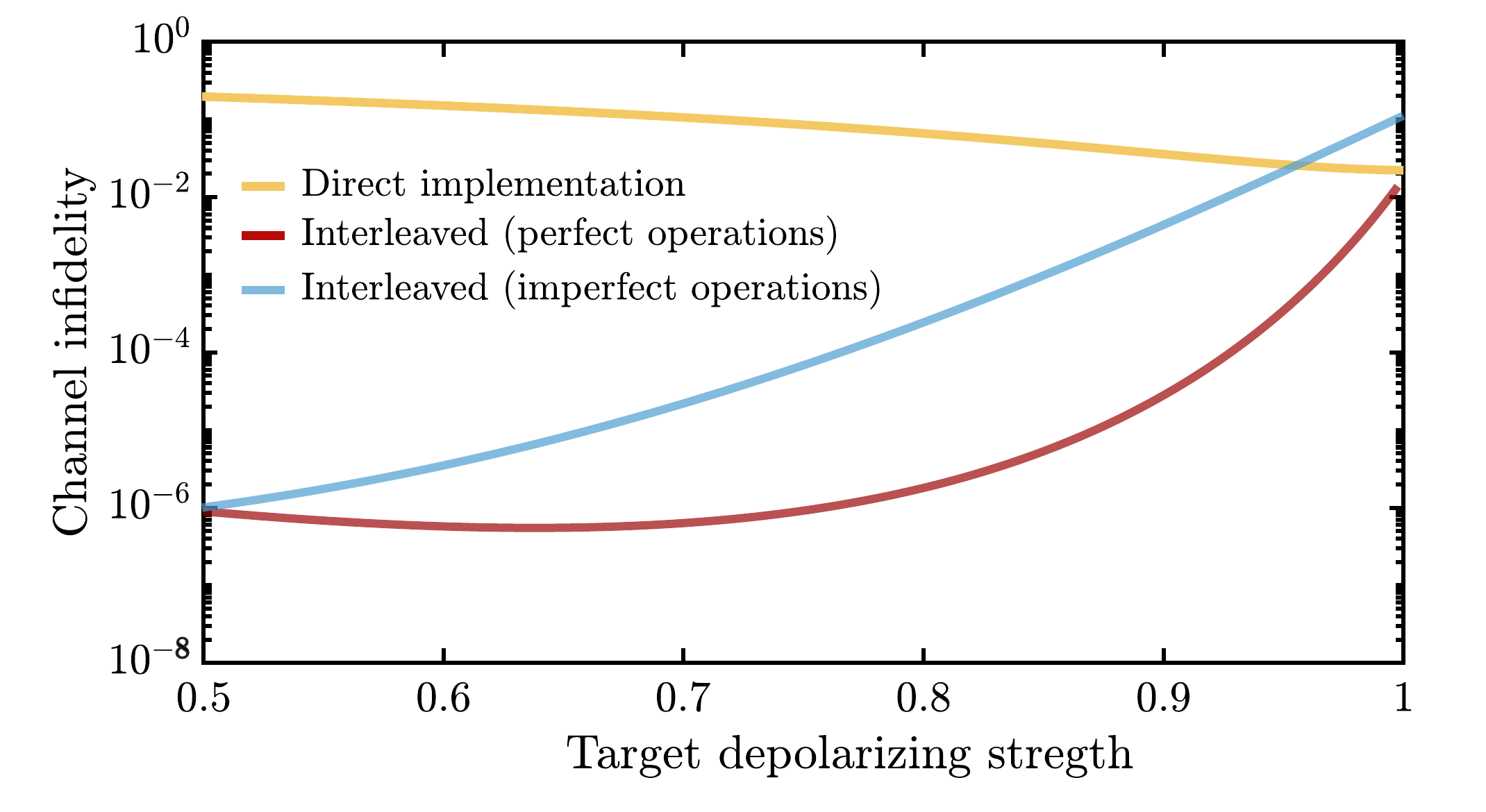} \label{fig:damp_to_depo_blocks}}
    \caption{\label{fig:blocks} (a) Channel infidelity $(1-F)$ for the simulation of a bit-flip channel (see also Appendix \ref{appendix.Bit.flip.channel}) with fixed parameter $p=0.95$ subjected to noise given by  $\hat{\mathcal{B}}$ (see text) with strength $q$ (x-axis), compared to the interleaved building-block strategy to compensate for the noise with perfect and imperfect hardware operation. The yellow line indicates the quality of direct channel simulation subjected to hardware noise, i.e., $\hat{\mathcal{E}}'_{\text{t}}$. (b)  Channel infidelity (Eq.~\eqref{eq:channel_fid}) when transforming an amplitude damping channel of $\gamma=0.1$, Eq.~\eqref{eq:dampingKrauses}, into a depolarizing channel of varying strength (x-axis) by using the interleaved strategy with perfect, and imperfect hardware, the latter modeled in a block fashion (Sec.~\ref{sec:back_model}) with depolarizing noise of strength $q=0.9$.}
\end{figure}

Fig.~\ref{fig:depo_blocks} presents an illustrative example where a bit-flip channel with $5\%$ noise strength is simulated using its basic circuit implementation (for details, refer to Appendix~\ref{appendix.Bit.flip.channel}). In this simulation, we model hardware noise following the block model discussed in Sec.~\ref{sec:back_model}, such that the circuit is ideally implemented, followed by the introduction of noise $\hat{\mathcal{B}}$ characterized by Kraus operators $\{K_0= q \, \id, K_i= \frac{1-q}{6} (\id + \ti \, \sigma_i) \}$, with the parameter $q$ varying along the x-axis. The choice of the noise is motivated given its non-Pauli diagonal nature, in order to illustrate how our method works in these scenarios too.

The objective here is to compensate for the inherent noise $\hat{\mathcal{B}}$ by employing the interleaved building block strategy. Each additional channel introduced to compensate for the noise is subjected to the same amount of noise identically modeled. The challenge lies in designing these additional channels in such a way that they effectively mitigate the inherent noise present in the simulation implementation. For comparative and fundamental purposes, we also analyze the scenario where the additional channels are not subjected to any noise. This example highlights the potential advantages of employing such an approach in specific regimes. Even when each block channel is subjected to the same level of noise, the interleaved strategy demonstrates its effectiveness in mitigating noise-induced errors. This observation underscores the potential utility of the approach in practical quantum simulation using NISQ devices scenarios. For a more in-depth analysis of bit-flip channel implementations, we direct the reader to Appendix~\ref{appendix.Bit.flip.channel}.

Instead of solely compensating for noise, this method can also be utilized for channel transformation, with the objective of converting some input channel $\hat{\mathcal{E}}{\text{in}}$ into a --in principle very different-- target channel $\hat{\mathcal{E}}{\text{t}}$ using the aforementioned techniques. In the context of large network simulations, there may arise scenarios where a specific type of noise needs to be simulated from a particular point. This strategy enables the direct transformation of the noise present in the simulation into the desired form, effectively harnessing the inherent noise of the quantum computer. From a fundamental perspective, the task of transforming one type of noise into another using additional channels applied before and after holds significant importance, even within the context of noiseless quantum hardware assumptions.

The problem of transforming channels into other channels is closely related to the concept of channel decomposition \cite{Wolf2008, Gong2023}. Channel decomposition analyzes how a given channel can be expressed as a composition of other channels. It has been demonstrated that any quantum channel can be decomposed into an arbitrary number of other channels \cite{Wolf2008}. In this context, our aim is to address this problem by employing a minimal number of additional channels, typically just two channels acting before, after, or both before and after the input channel. This pragmatic approach allows us to achieve channel transformation with minimal resources. For example, transforming a bit-flip channel into a dephasing channel and vice versa can be achieved by simply applying a Hadamard unital channel before and after the input channel (up to inherent noise). However, we also consider scenarios involving highly non-trivial transformations, particularly those between unital and non-unital channels.

Fig.~\ref{fig:damp_to_depo_blocks} illustrates the effectiveness of this approach in transforming an amplitude-damping channel into a depolarizing channel (see Sec.~\ref{sec:back_model}). We present results obtained under both noiseless and noisy quantum computer operations. The noise is again modeled following the block model described in Sec.~\ref{sec:back_model}, which consists of depolarizing noise with fixed strengths of $10\%$. Remarkably, the ability to apply block channels in an interleaved fashion demonstrates significant tunability of the different channels across most regimes. This holds true even when the additional channels are subjected to noise, highlighting the robustness and versatility of the approach.

Further extensions, for instance, considering a larger number of concatenated channels (before and after), are expected to further enhance the accuracy of these strategies.

\subsection{Method 2. Tailored circuit channels} \label{sec: Method 4}

In the context of simulating quantum channels within circuits, it is natural to explore whether manipulating gate sequences or adjusting gate parameters could lead to more accurate channel transformations, either to harness and convert or to directly compensate for the inherent noise of the quantum computer. In this method, we propose a flexible strategy where the circuits simulating the desired channels are not rigidly defined. Instead, one has the freedom to shape these circuits to effectively incorporate the hardware noise and better approximate the desired channel implementations, see Fig.~\ref{fig:enter-label}(b).

Through relevant analytical and numerical examples, we illustrate the efficacy of this approach. Analytically, we explore various possibilities considering different hardware noise assumptions, where we demonstrate how varying specific parameters within the quantum gates that implement these channels, is possible to compensate for inherent noise and significantly improve simulation accuracy. Numerically, we further investigate this method by considering two distinct scenarios. In the first scenario, we allow for complete flexibility in shaping the circuits, enabling adjustments to all gate parameters. In contrast, the second scenario imposes constraints, permitting only certain parameters of the gates to be tuned. By comparing the results of these numerical analyses, we gain valuable insights into the practical implications and benefits of this tailored circuit strategy.

\textit{Analytical example 1. Pauli-diagonal noise.--} The first scenario of interest involves quantum hardware with noise modeled according to the gate model outlined in Sec.~\ref{sec:back_model}, specifically with Pauli diagonal noise. In this model, the implementation of a unitary $U$ gate is represented as  $U \rho \, U^\dagger \; \mapsto \; \hat{\mathcal{E}} \left( U \rho \, U^\dagger \right)$, where
\begin{equation}
    \label{eq:pauli.diagonal.noise}
    \hat{\mathcal{E}} (\rho) = \sum_{i\in |\textbf{P}_n|} p_i \, \Sigma_i \, \rho \, \Sigma_i,
\end{equation}
where $\Sigma_i$ is an element of the $n$-Pauli group, $\textbf{P}_n$. 

Any Pauli diagonal channel of the form $\hat{\mathcal{E}}$ can be transformed into a different Pauli diagonal channel by implementing specific post-processing operations. This transformation is achieved by applying imperfect (note these gates are also subjected to Pauli diagonal noise) Pauli gates $\hat{\mathcal{E}}(\Sigma_i \bullet \Sigma_i)$ with probability $\lambda_i$ after~$\hat{\mathcal{E}}$. Mathematically, this transformation can be expressed as
\begin{align*}
    \hat{\mathcal{E}}(\rho) & \to \sum_{i,j\in |\textbf{P}_n|} \lambda_j \, \hat{\mathcal{E}} \left[ \Sigma_j \, \mathcal{\hat{E}}( \rho ) \, \Sigma_j \right] \\ 
    & = \sum_{i,j,k\in |\textbf{P}_n|} q_i \, \lambda_j \, p_k \Sigma_i \Sigma_j \Sigma_k \, \rho \, \Sigma_k\Sigma_j\Sigma_i \\ 
    & = \sum_{i\in |\textbf{P}_n|} \tilde{p}_i \, \Sigma_i \, \rho \, \Sigma_i,
\end{align*}
where $\{q_i\}$ is another probability distribution. The resulting channel remains $n$-Pauli diagonal with coefficients ${ \tilde{p}_i }$, the process can be adapted, through appropriate selection of ${ \lambda_i }$, to achieve the desired channel transformation.

In the particular case where the hardware Pauli diagonal noise corresponds to single qubit depolarizing noise, i.e., $\hat{\mathcal{E}} = \hat{\mathcal{D}}$, we observe that $\tilde{p}_i = PQ \, \lambda_i + (1-QP)/4$, where $P = (4p_0-1)/3$ and $Q = (4q_0-1)/3$. Consequently, for a target Pauli diagonal channel characterized by $\{ \tilde{p}_i \}$, given that it satisfies the condition:
\begin{equation}
    \frac{1-PQ}{4} \leq \tilde{p}_i \leq PQ + \frac{1-PQ}{4},
\end{equation}
we can effectively simulate it by setting $\lambda_i = (4 \tilde{p}_i + PQ -1)/(4PQ)$.

\textit{Analytical example 2. Amplitude damping noise.-- } We delve here into interesting tunability properties of the amplitude damping channel (refer also to Appendix~\ref{appendix.Alternative.circuits.for.amplitude.damping.channel}). In this scenario, the amplitude damping channel serves as the desired simulation implementation, and also as a model for the inherent hardware noise.

Specifically, we observe that the decomposition of two amplitude damping channels with parameters $P_1$ and $P_2$ results in yet another amplitude damping channel, with parameter $\tilde{P}=P_1+P_2-P_1P_2$. Consequently, by applying a given noisy operation described by amplitude damping multiple times, one can simulate amplitude damping channels with higher $\tilde{P}$. It is worth noting, however, that only discrete values are attainable (more if $P$ is small), and $\tilde P \rightarrow 1$ for multiple applications. For $n$ applications, the effective noise parameter $\tilde P_n$ can be described by a specific polynomial of degree $n$. This already suffices to obtain excellent approximations of the desired noise process, with a fidelity $F_1$. 

A situation that cannot be nicely handled with this approach is cases where the achievable discrete values are far from the desired one, e.g., if $\tilde P$ is only slightly larger than $P$. In these cases, an implementation as indicated in Fig.~\ref{fig.ampitude.damping.a}, i.e., consisting of a controlled rotation gate into an ancilla followed by a cNOT gate and a tracing out of the ancilla, can achieve higher accuracy, by properly tuning $\theta$. Notice that choosing $\sin(\theta/2)=\sqrt{\tilde P}$ gives an exact simulation of the desired amplitude damping channel in the noiseless case. However, using this value when the controlled-$R_y(\theta)$ rotation is itself noisy (described by amplitude damping with parameter $P$ acing on both output qubits) is on the one hand no longer an amplitude damping channel. On the other hand, the channel fidelity $F_2$ is pretty low. One can optimize $\theta$ though, and obtain a fidelity $F_3$. In most cases (small $P$, and $\tilde P$ much larger - about an order of magnitude or more) $F_1 > F_3 > F_2$. However, e.g., for $\tilde P=0.45$, $P=0.4$ one obtains $F_3 > F_1 > F_2$.  

\begin{figure}
    \centering
    \subfloat[]{\includegraphics[width=0.9\columnwidth]{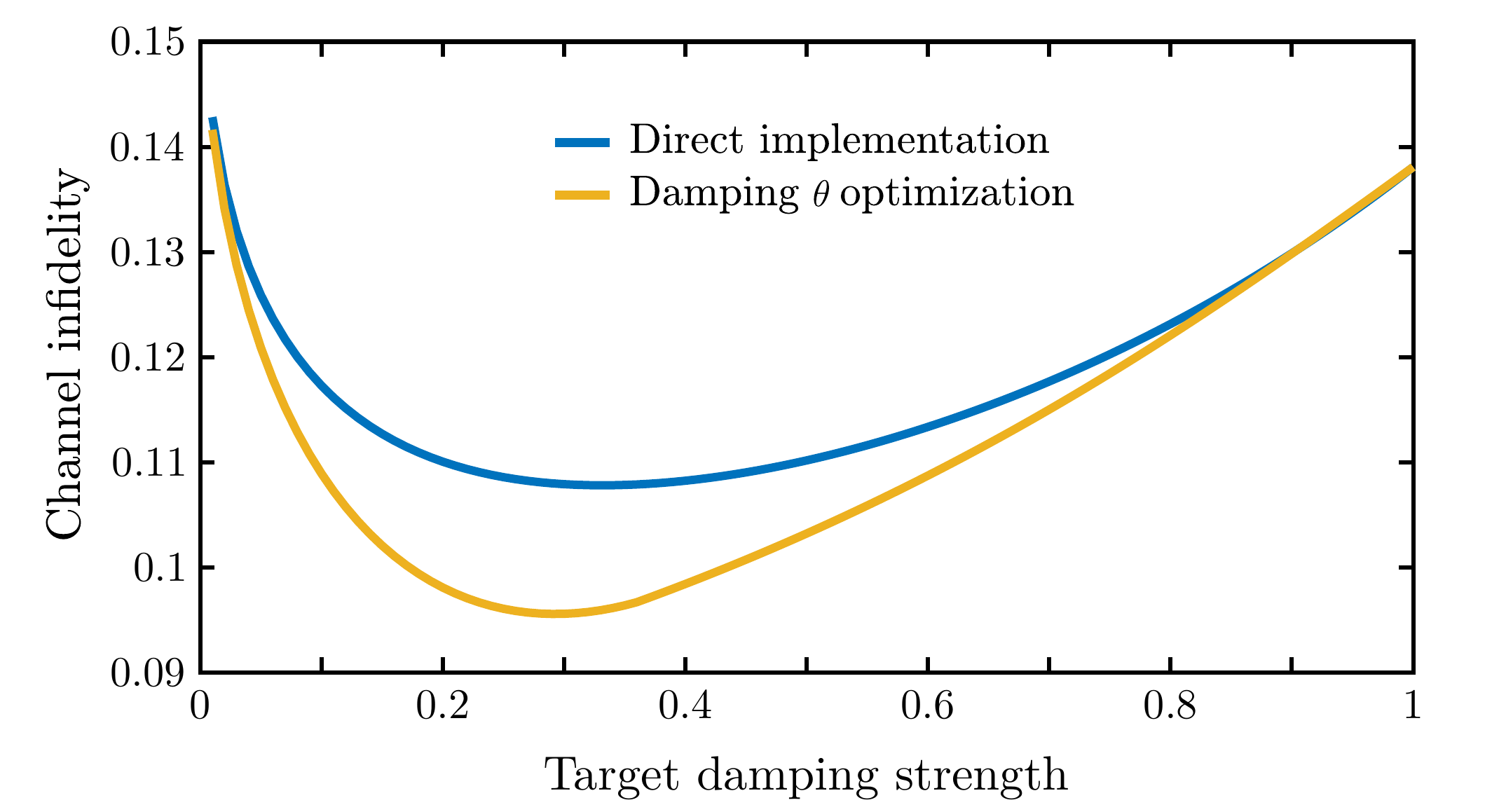} \label{fig:damp_gates}} \\
    \subfloat[]{\includegraphics[width=0.9\columnwidth]{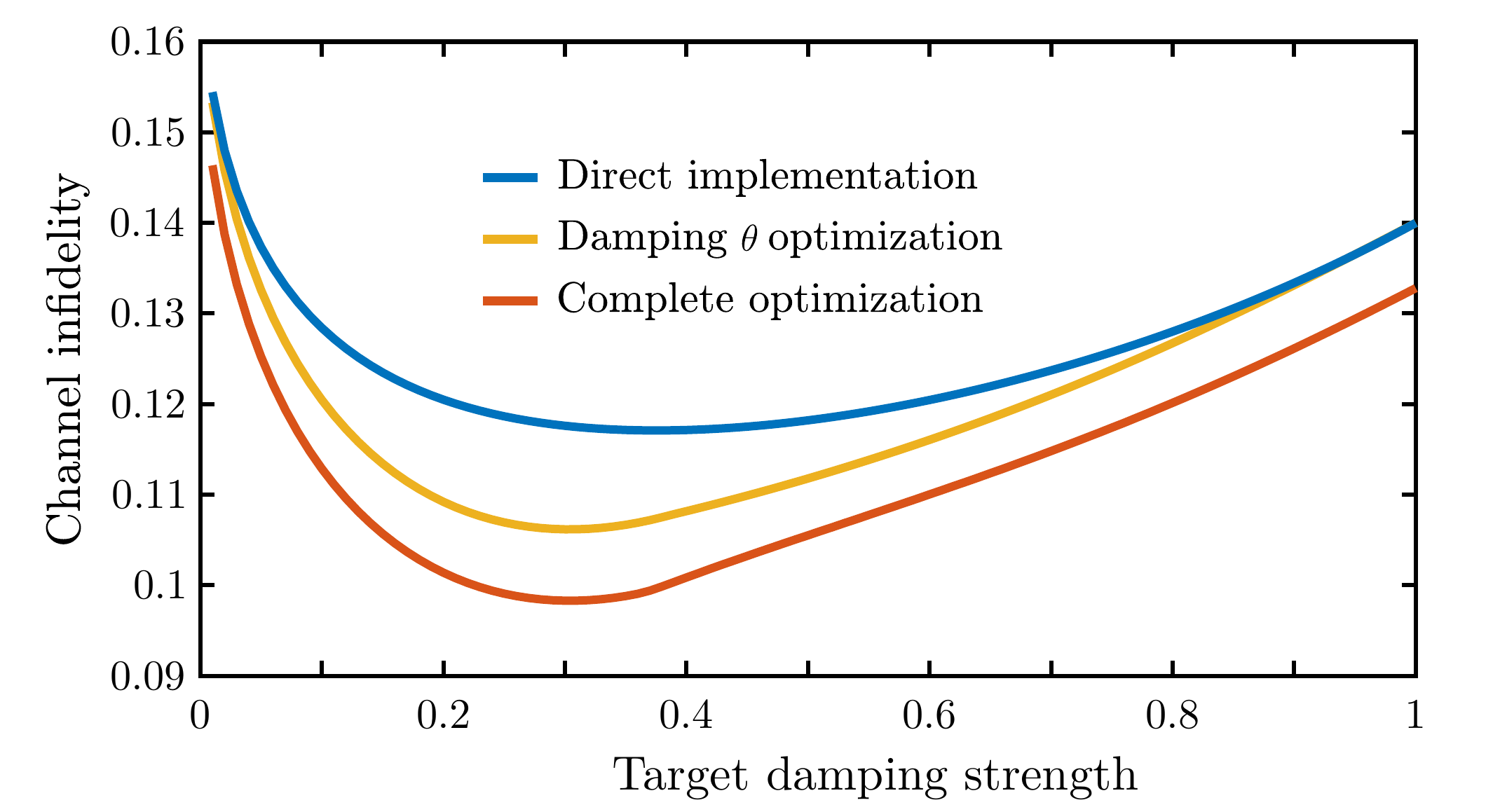} \label{fig:damp_block}} \\
    \subfloat[]{\includegraphics[width=0.9\columnwidth]{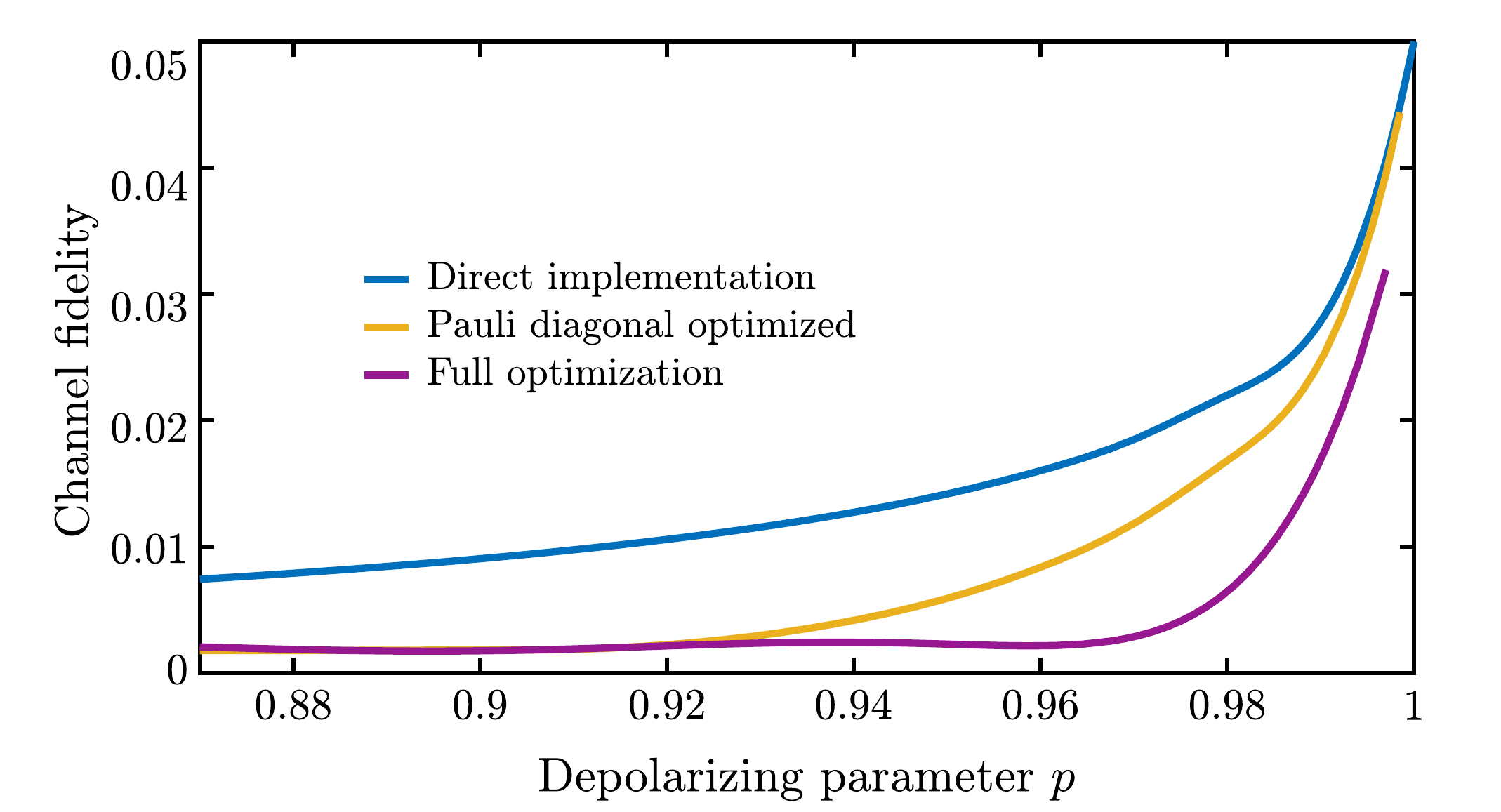} \label{fig:regimes}}
    \caption{\label{fig:depotodamp} (a) Amplitude damping channel simulation with target strength $\gamma$ given by the x-axis. Noise is modeled in a noisy-gate fashion, see Sec.~\ref{sec:back_model}, where all the gates are affected by depolarizing and dephasing noise with parameter $q=0.925$ in each register. (b) Amplitude damping channel implementation with hardware noise following block model. All the channels are affected by depolarizing and dephasing noise of parameter $q=0.8$ each. (c) Simulation of a depolarizing channel subjected to amplitude damping and dephasing hardware noise of parameter $q=0.8$.}
\end{figure}

\textit{Numerical analysis.-- } Fig.~\ref{fig:depotodamp} illustrates the advantages of employing the tailored circuit strategy for simulating different channels in a quantum device across various noise scenarios.

In Fig.~\ref{fig:damp_gates}, we observe the performance of implementing an amplitude damping channel, as described by Eq.~\eqref{eq:dampingKrauses}, with varying strength $\gamma$ (x-axis). The simulation is conducted using the basic circuit depicted in Fig.~\ref{fig.ampitude.damping.b}, where hardware noise is modeled in a noisy-gate fashion, following the approach outlined in Sec.~\ref{sec:back_model}. Specifically, noisy channels act in each register after every gate, as illustrated in Fig.~\ref{fig.ampitude.damping.c}. Here, the hardware noise is fixed as depolarizing followed by dephasing, with a strength of $7.5\%$ each. Adopting the tailored circuit strategy, we focus on tuning only the parameter $\theta$ from the gate of the circuit in Fig.~\ref{fig.ampitude.damping.c}. The figure demonstrates how the optimal controlled-rotation parameter $\theta$, where $\sin^2\!\left({\theta/2}\right) = \gamma$ varies depending on the computer noise. This observation underscores how optimizing this parameter can enhance the quality of channel simulation.

In addition, Fig.~\ref{fig:damp_block} explores a similar approach but with noise modeled according to the block model outlined in Sec.~\ref{sec:back_model}, again consisting of depolarizing followed by dephasing, but with a strength of $20\%$ each. In this scenario, we also allow for full circuit tunability, and the results indicate that this approach is generally even more beneficial. Notice that channel infidelities are significantly high due to the substantial amount of inherent noise assumed in these simulations.

Finally, in Fig.~\ref{fig:regimes}, assuming again the block noise model now with dephasing followed by amplitude damping noise, the objective is the simulation of a depolarizing channel. We consider different tunability variants: the direct implementation, entailing the application of each Pauli operator with equal probability; Pauli diagonal optimization, enabling optimization over the probabilities associated with each Pauli operator; and full circuit optimization. Notably, even partial optimization, specifically tuning the probabilities of Pauli operators, yields significant enhancements compared to direct channel simulation. These enhancements can be further augmented when granting full flexibility in circuit design.

These examples underscore the potential of leveraging quantum computers for simulating quantum communication processes. Importantly, advantages are discerned across all cases, irrespective of the noise model under consideration, although significantly better accuracy is found when dealing with similar kinds of noise. Furthermore, it is noteworthy that similar qualitative behavior persists across diverse hardware noise sources and analyzed regimes.
 %

%
%
\subsection{Method 3.  Variational black-box optimization} \label{sec:Variational}
In this section we discuss a versatile method that complements the other proposed approaches and strategies, offering an alternative perspective on circuit simulation. The core concept involves treating the circuit simulation as a black-box optimization problem, where only certain parameters are accessible for adjustment. The simulation runs physically on a quantum device, iteratively updating input parameters to optimize an output figure of merit, such as the fidelity of the simulation compared to the desired outcome. This approach draws parallels to Variational Quantum Optimization (VQO) techniques \cite{Doolittle2023, qNetVO, Doolittle2024}., enabling classical optimization of the effective channel implemented in the quantum computer by leveraging a chosen figure of merit to guide parameter updates across different circuit iterations. Recent works have also combined this idea with variational optimization to solve or optimize network problems under noise \cite{Doolittle2023, qNetVO, Doolittle2024}. In contrast, our work adopts a complementary viewpoint: rather than optimizing the network using quantum hardware, we focus on tailoring the simulation itself—adapting it to the device noise features to better reproduce network processes with high accuracy and in a noise-blind way.

Note that the specific figure of merit and the parameters to be optimized are problem-dependent and can vary with the structure and context of the simulated setting.

A particularly advantageous aspect of this variational approach lies in its practicality. Firstly, it is hardware-independent, eliminating the need for prior knowledge about the inherent noise characteristics of the device. It operates in a black-box simulation fashion, where users iteratively optimize over classical parameters fed into the quantum device, which, in turn, provides information about the figure of merit in such black-box manner.

To illustrate this variational strategy further, consider Method 2 (described in Sec.~\ref{sec: Method 4}), applied for simulating an amplitude-damping channel. The quantum device tasked with implementing the corresponding circuit (depicted Sec.~\ref{appendix.Alternative.circuits.for.amplitude.damping.channel} can be viewed as a black box, where the parameter $\theta$ of the controlled rotation $R_y(\theta)$ can be freely adjusted. Optimization over this parameter, using as figure of merit the channel Choi fidelity, Eq. (\ref{eq:channel_fid}), compared to the target channel, maximizes the quality of the simulation without necessitating knowledge of the inherent noise characteristics of the device.

Similar techniques can be employed with different input parameters or optimized over arbitrary circuit implementations of channels (or network processes in general). This flexibility furnishes a potent and general simulation tool capable of operating independently of specific inherent noise sources in quantum devices.

\subsection{Demonstration in real hardware}
We demonstrate the practical feasibility of our approach by implementing the basic techniques directly on quantum hardware. This serves as a proof-of-concept for the possibility of tailoring the inherent noise of a quantum computer to engineer specific quantum channels, and eventually whole communication simulations.

\begin{figure}[h!]
    \centering
    {\includegraphics[width=0.9\columnwidth]{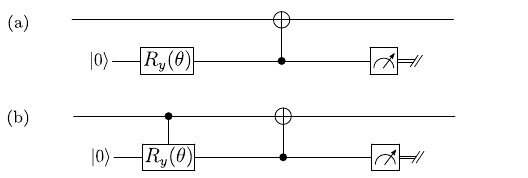}} 
    \caption{\label{fig:fig_real_hardware.circuit} Circuit implementation for real hardware reproduction of (a) Bit-flip channel with strength parameter given by $1-p= \sin^2(\theta/2)$, and $R_y = \text{exp}\{-i \frac{\theta}{2} Y\}$. (b) Amplitude damping channel with decay parameter $\gamma = \sin^2(\theta/2)$. (see also Appendices \ref{appendix.Bit.flip.channel}, \ref{appendix.Alternative.circuits.for.amplitude.damping.channel})}
\end{figure}

\begin{figure}[h!]
    \centering
    {\includegraphics[width=0.9\columnwidth]{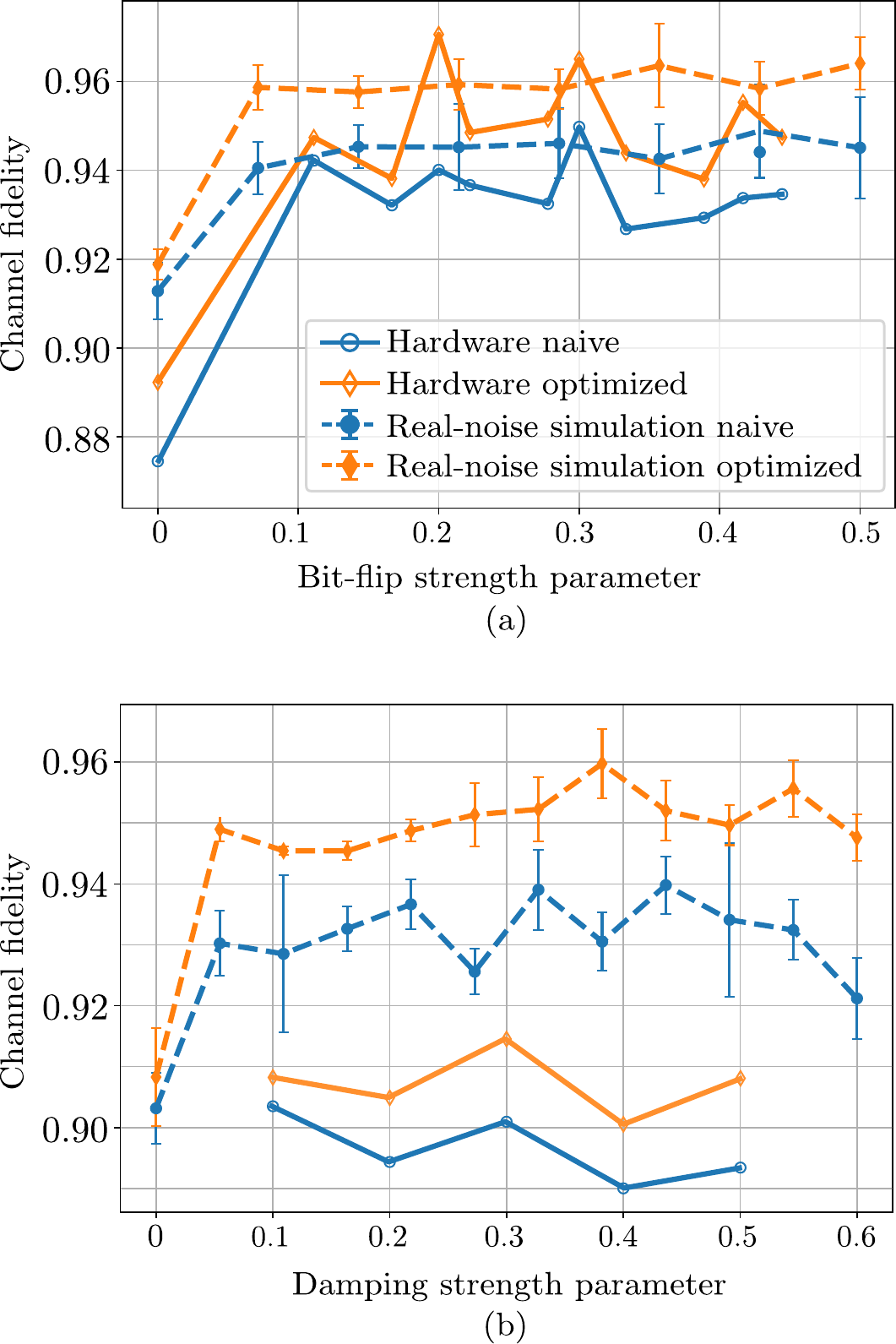}} 
    \caption{\label{fig:Figure_realhardware} Demonstration on real hardware using the IBM Quantum Platform  \textit{ibm\_torino} device. Realistic backend calibration data is used for the simulation (dashed) curves and real hardware for solid curves. Theoretical best (naive) and variationally tailored (optimized) rotation parameters are considered. Tomographic process reconstruction over $2048$ shots is used in all cases, where channel (Choi) fidelity, compared to the ideal channel implementation, is computed from them. (a) Bit flip channel. (b) Amplitude damping channel.}
\end{figure}

Specifically, we realize the circuits shown in Figs.~\ref{fig:fig_real_hardware.circuit} (a) and (b), which correspond to ancilla-assisted simulations of a bit-flip channel and an amplitude damping channel, respectively (see also Secs. \ref{appendix.Bit.flip.channel}, \ref{appendix.Alternative.circuits.for.amplitude.damping.channel}). For each case, we compare the naive (theoretically optimal in the absence of noise) parameterization of the rotation angle to a version in which this parameter is variationally optimized in the presence of hardware noise, see Fig.~\ref{fig:Figure_realhardware}. In both cases, we observe that fine-tuning the circuit parameters leads to a better approximation of the target quantum channel.

Experiments were carried out on the IBM Quantum Platform, using the \textit{ibm\_torino} device. For simulation results, we replicated the noise model as faithfully as possible using the real hardware calibration data. The experimental implementations were performed on the same backend for consistency.


Our results validate the proposed approach, i.e., even a simple optimization of the rotation parameter results in a significant improvement in channel reproduction fidelity. We note that the overall lower fidelities observed for the amplitude damping in the real hardware case are due to the non-trivial decomposition of the controlled-Y rotation in the machine, which increases circuit depth and cumulative noise. Nevertheless, the optimized case consistently outperforms the naive approach,  highlighting the benefit of noise tailoring on real devices.

\section{Conclusions} \label{sec:conclusions}





In this study, we have presented an alternative strategy for simulating quantum network processes using quantum hardware, with the goal of complementing existing classical simulation methods, which remain essential tools in the field and have seen considerable progress in recent years \cite{Coopmans_2021, Wu_2021, Satoh_2022, Julius2022, Chen_2023, Bassoli2020}.

Classical simulators are extremely valuable, especially in regimes where efficient techniques such as stabilizer methods apply. In fact, much of the progress in network modeling has been achieved by focusing on classically accessible and simulable scenarios. However, they are fundamentally limited in situations involving large-scale networks or general noise models beyond the Clifford or Pauli framework, where the exponential scaling of resources becomes prohibitive. In contrast, quantum computers offer a natural platform for reliable communication simulation in such regimes, as they operate directly in the same Hilbert space we aim to explore.

Our contribution lies in identifying and addressing the key technical ingredients that enable accurate and practical simulations using quantum hardware, including noise tailoring, protocol design, and characterization strategies. Our approach takes a different perspective, rather than mitigating the noise inherent in NISQ devices, we exploit and reshape it to match the noise characteristics of realistic quantum memories, gates, and communication channels. We have introduced tools to tailor local quantum noise, achieving high-fidelity simulations of target channels and demonstrating (experimentally on real hardware) how this strategy enables the simulation of complete communication protocols on current quantum hardware.

Our method still presents limitations that require further investigation. Apart from the natural limitations of current quantum hardware, the tailoring techniques are currently restricted to single-qubit noise, and while we achieve good accuracy when matching similar noise types, extending to multi-qubit or improving the accuracy for more general noise, or with additional tools, remains an open challenge. Additionally, although circuit depths remain relatively constant when simulating completely positive maps, measurement overheads can be significant. Exploring how to reduce the need for full process tomography --for instance evaluating the Choi matrix directly-- is thus an important question for future work.

We emphasize that the limitations we face are different from those of classical simulators. The boundaries of quantum hardware-based simulation are not yet well understood, and future research will be essential in clarifying what is possible. Still, our work aims to lay the foundation for a complementary approach to network simulation, offering new opportunities particularly in regimes where classical methods struggle.


\section*{Acknowledgments}
This research was funded in whole or in part by the Austrian Science Fund (FWF) 10.55776/P36009  and 10.55776/P36010. For open access purposes, the author has applied a CC BY public copyright license to any author-accepted manuscript version arising from this submission. Finanziert von der Europäischen Union - NextGenerationEU.

We acknowledge support by the Austrian Research Promotion Agency (FFG)
under Contract Number FO999911051/52065529.

We acknowledge the use of IBM Quantum Platform services for this work.

\bibliography{QQsimulator.bib}

\appendix

\section{Channel dilatation with an ancillary system}
\label{appendix.dilatation.1}
Stinespring dilation theorem shows that given a quantum channel $\hat{\mathcal{E}}$ one can always find a unitary gate $\Lambda: \mathcal{L}(\mathcal{H}_{\bar{A}} \otimes \mathcal{H}_A) \to \mathcal{L}(\mathcal{H}_{\bar{B}} \otimes \mathcal{H}_B)$ such that
\begin{equation}
    \label{eq.appen.dilatation1}
    \hat{\mathcal{E}}(\rho) = \text{tr}_{\bar{B}} \!\left[ \Lambda \left( \proj{0}_{\bar{A}} \otimes \rho \right) \Lambda^\dagger \right],
\end{equation}
where $dim(\mathcal{H}_{\bar{A}} \otimes \mathcal{H}_A) = dim(\mathcal{H}_{\bar{B}} \otimes \mathcal{H}_B)$.

In this section, we consider the case where $\mathcal{H}_A = \mathcal{H}_B$ and $\mathcal{H}_{\bar{A}} = \mathcal{H}_{\bar{B}}$, and we show that if $dim(\mathcal{H}_{\bar{A}}) = r$ where $r$ is the Kraus rank of $\hat{\mathcal{E}}$, then $\Lambda$ always exist.

Being $\{ K_i \}_{i=0}^{r-1}$ a minimal Kraus representation of $\hat{\mathcal{E}}$, if $\Lambda$ fulfils
\begin{equation}
\label{eq.appen.condition1}
    \bra{i, k} \Lambda \ket{0, l} = \bra{k} K_i \ket{l},
\end{equation}
or equivalently in the computational basis, it is of the form 
\begin{equation}
    \label{eq:condition1.2}
    \Lambda =
    \begin{pmatrix}
        \;\boxed{\;\; K_0 \;\;}\; & \bullet & \bullet & \bullet \;\; \\[5pt]
        \;\boxed{\;\; K_1 \;\;}\; & \bullet & \bullet & \bullet \;\; \\
        \vdots & & &                                                          \\[3pt]
        \;\boxed{K_{r-1}}\;       & \bullet & \bullet & \bullet \;\;
    \end{pmatrix},
\end{equation}
then its action on the system is given by
\begin{equation}
\begin{aligned}
    \Lambda \left( \proj{0} \otimes \rho \right) \Lambda^\dagger = \sum_{i,j=0}^{d-1} \rho_{ij} \Lambda \ketbra{0,i}{0,j} \Lambda^\dagger \\
    = \!\!\sum_{\substack{0\leq l,i,j,n<d \\ 0\leq k,m < r}} \! \rho_{ij} \bra{k,l}\Lambda\ket{0,i}\bra{0,j} \Lambda^\dagger\ket{m,n} \ketbra{k,l}{m,n} \\
    \underset{\eqref{eq.appen.condition1}}{=} \sum_{\substack{0\leq l,i,j,n<d \\ 0\leq k,m < r}} \rho_{ij} \bra{l} K_k \ket{i} \bra{j}K_m^\dagger\ket{n} \ketbra{k,l}{m,n} \\
    = \sum_{\substack{0\leq l,n<d \\ 0\leq k,m < r}}  \ketbra{k}{m} \otimes \proj{l} K_k \, \rho \, K_m^\dagger \proj{n} \\
    = \sum_{k,m =0}^{r-1}  \ketbra{k}{m} \otimes K_k \, \rho \, K_m^\dagger,
\end{aligned}
\end{equation}
where $\rho_{ij} = \bra{i}\rho\ket{j}$.

Then, note that if the ancilla state is traced out the quantum channel is implemented to the state $\rho$, i.e.,
\begin{equation}
\begin{aligned}
    \text{tr}_{\bar{B}} \!\left[ \Lambda \left( \proj{0}_{\bar{A}} \otimes \rho \right) \Lambda^\dagger \right] & = \sum_{k,m =0}^{r-1} \bracket{m}{k} \otimes K_k \, \rho \, K_m^\dagger \\
    & = \sum_{k=0}^{r-1} K_k \, \rho \, K_k^\dagger = \mathcal{E}(\rho),
\end{aligned}
\end{equation}

Therefore, condition Eq.~\eqref{eq.appen.condition1} guarantees that $\Lambda$ is a dilation of channel $\hat{\mathcal{E}}$. Now, we still need to show that the condition in Eq.~\eqref{eq.appen.condition1} is always compatible with $\Lambda$ being unitary.

$\Lambda$ is a unitary matrix if and only if it can be written as
\begin{equation}
    \Lambda = \sum_{\substack{0\leq k < r \\ 0 \leq l < d}} \ketbra{A_{kl}}{k,l},
\end{equation}
where $\{ \ket{A_{kl}}\}$ is an orthonormal basis. Condition in Eq.~\eqref{eq.appen.condition1} completely determines $\{ \ket{A_{0l}} \}_{l=0}^{d-1}$ which are given by
\begin{equation}
    \ket{A_{0l}} = \sum_{\substack{0 \leq i < r \\ 0 \leq j < d}} \bra{j} K_i \ket{l} \ket{i,j}.
\end{equation}
Note that by construction it is fulfilled that
\begin{equation}
    \bracket{A_{0k}}{A_{0l}} = \sum_{\substack{0 \leq i < r \\ 0 \leq j < d}} \bra{k} K^\dagger_i \ket{j} \bra{j} K_i \ket{l} = \delta_{kl},
\end{equation}
where we have used that $\sum_{i=0}^{r-1} K_i^\dagger K_i = \id$. Therefore, $\{ \ket{A_{0l}} \}$ form an orthonormal set. Then, we always can find the rest of vectors $\{ \ket{A_{i0}}, \dots, \ket{A_{i,d-1}}\}_{i=0}^{r-1}$ such that $\{ \ket{A_{ij}}\}$ forms an orthonormal basis, i.e., completing matrix $\Lambda$ such as it is unitary.

\section{Channel dilatation with an extended qudit}
\label{appendix.dilatation.D}
In this section, we consider the super-qudit channel implementation, where a qudit state of dimension $d$ is encoded in the so-called \textit{data subspace} given by $span\{\ket{k}\}_{k=0}^{d-1}$, where the total of a Hilbert space is of dimension $D$. Given a quantum channel $\hat{\mathcal{E}}$, our goal is to find a routine $\hat{\mathcal{P}}$ consisting of unitary gates and projective measurements acting on the super-qudit such that implements the quantum channel on the qudit state, i.e.,
\begin{equation}
    \hat{\mathcal{P}}(\rho \oplus \bo{0}) = \hat{\mathcal{E}}(\rho) \oplus \bo{0}.
\end{equation}

Being $\{ K_i \}_{i=0}^{r-1}$ a minimal Kraus representation of channel $\hat{\mathcal{E}}$, we first perform the single value decomposition of $K_i$, and write the Kraus operators as $K_i = W_i \Sigma_i V_i^\dagger$. Without loss of generality, we find $W_i$ and $V_i$ such that $\Sigma$ fulfils $\Sigma_{ij} \geq \Sigma_{i,j+1}$. With this choice, we define $\tilde{K}_i$ as a matrix consisting of the non zero rows of $\Sigma_i V_i^\dagger$, i.e.,
\begin{equation}
\label{eq:G}
    \Sigma_i V_i^\dagger =
    \begin{pmatrix}
    \;\boxed{\; \tilde{K}_i \;}\;\; \\[5pt] 
    \;\boxed{\;\; \bo{0} \;\;} \;\;
    \end{pmatrix},
\end{equation}
where $dim(\tilde{K}_i) = \kappa_i \times d$, and note they fulfill
\begin{equation}
    \label{eq_GGI}
    \sum_{k=0}^{r-1} \tilde{K}_k^\dagger \tilde{K}_k = \id_d.
\end{equation}

Then, we construct a unitary matrix of the form
\begin{equation}
    \label{eq:condition1.D}
    \Lambda =
    \begin{pmatrix}
        \;\boxed{\;\; \tilde{K}_0 \;\;}\; & \bullet & \bullet & \bullet \;\; \\[5pt]
        \;\boxed{\;\; \tilde{K}_1 \;\;}\; & \bullet & \bullet & \bullet \;\; \\
        \vdots & & &                                                          \\[3pt]
        \;\boxed{\tilde{K}_{r-1}}\;       & \bullet & \bullet & \bullet \;\;
    \end{pmatrix},
\end{equation}
where $\Lambda$ is a unitary gate acting on a Hilbert space of dimension $D = \sum_{i=0}^{r-1} \kappa_i$. Applying $\Lambda$ to the super-qudit we obtain
\begin{equation}
\label{eq.Lambda.D}
    \Lambda \left( \rho \oplus \bo{0} \right) \Lambda^\dagger = 
    \begin{pmatrix}
        \,\boxed{\tilde{K}_0 \, \rho \, \tilde{K}_0^\dagger} & \,\boxed{\tilde{K}_0 \, \rho \, \tilde{K}_1^\dagger} & \cdots \\[6pt]
        \,\boxed{\tilde{K}_1 \, \rho \, \tilde{K}_0^\dagger} & \,\boxed{\tilde{K}_1 \, \rho \, \tilde{K}_1^\dagger} & \cdots \\
        \,\vdots & \vdots \\
    \end{pmatrix}.
\end{equation}
Then we perform a projective measurement of the super-qudit such that projects to the subspaces corresponding to the diagonal blocks given by $\tilde{K}_i \rho \, \tilde{K}_i^\dagger$, i.e., the projective measurement is given by $\{ P_i \}_{i=0}^{r-1}$ where
\begin{equation}
    P_i = \sum_{j= c_{i-1} }^{ c_{i} -1} \proj{j},
\end{equation}
with $c_i = \sum_{j=0}^i \kappa_j$.

Next, we perform a correction operation $\bar{W}_i= (W_i \oplus \bo{0}) X_D^{c_{i-1}}$, where $X_D \ket{j} = \ket{(j-1)\text{mod} \, D}$, which first moves the qudit state into the data subspace and then applies $W_i$ implementing $p^{-1}_i \, K_i \, \rho \, K_i^\dagger$, where $p_i = \text{tr} \!\left[P_i \Lambda (\rho \oplus \bo{0} ) \Lambda^\dagger\right]$. Finally, the outcome of the measurement is ``erased'' leading to the implementation of the channel, i.e.,
\begin{equation}
    \hat{\mathcal{P}}(\rho \oplus \bo{0}) = \sum_{i=0}^{r-1} \bar{W}_i \, P_i \, \Lambda \left( \rho \oplus \bo{0} \right) \Lambda^\dagger \, P_i \, \bar{W}^\dagger_i = \hat{\mathcal{E}}(\rho)\oplus \bo{0}.
\end{equation}
Therefore the detailed procedure implements the quantum channel to the qudit state. However, we still need to show that one always can find a unitary matrix $\Lambda$ that fulfills Eq.~\eqref{eq:condition1.D}.

We make use of the properties that $\Lambda$ is a unitary matrix if and only if it can be written as
\begin{equation}
    \Lambda = \sum_{k=0}^{D-1} \ketbra{A_k}{k},
\end{equation}
where $\{\ket{A_k}\}$ is an orthonormal basis. As $\ket{A_k}$ corresponds to the $k$th raw of $\Lambda$ we have that
\begin{equation}
    \ket{A_k} = \bigoplus_{i=0}^{r-1} \sum_{l_i=0}^{\kappa_i-1} \bra{l_i} \tilde{K}_i \ket{k} \ket{l_i} \quad \text{for } 0 \leq k < d
\end{equation}
Note that, by construction, it is fulfilled that
\begin{equation}
\begin{aligned}
    \bracket{A_m}{A_n} & = \sum_{i=0}^{r-1} \sum_{k_i,l_i=0}^{\kappa_i-1} \bra{m} \tilde{K}_i^\dagger \ket{k_i} \bra{l_i} \tilde{K}_i \ket{n} \bracket{k_i}{l_i} \\
    & = \sum_{i=0}^{r-1} \sum_{k_i=0}^{\kappa_i-1} \bra{m} \tilde{K}_i^\dagger \ket{k_i} \bra{k_i} \tilde{K}_i \ket{n} \\
    & = \sum_{i=0}^{r-1}  \bra{m} \tilde{K}_i^\dagger \; \underset{\id_{\kappa_i}}{\underbrace{\sum_{k_i=0}^{\kappa_i-1} \proj{k_i}}} \; \tilde{K}_i \ket{n} \\
    & = \sum_{i=0}^{r-1} \bra{m} \tilde{K}_i^\dagger \tilde{K}_i \ket{n} \underset{\eqref{eq_GGI}}{=} \delta_{mn}
\end{aligned}
\end{equation}
for $0 \leq m,n <d$.

Therefore, as there is no restriction about the rest of the matrix, vectors (columns) $\{ \ket{A_k} \}_{k=d}^{D-1}$ can be freely chosen such that $\{ \ket{A_k} \}_{k=0}^{D-1}$ form an orthonormal set, meaning $\Lambda$ is unitary.

\section{Bit flip channel}
\label{appendix.Bit.flip.channel}
\begin{figure}
    \centering
    \subfloat[]{\includegraphics[width=\columnwidth]{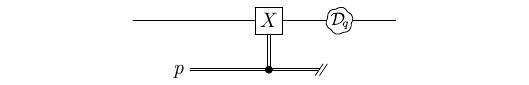} \label{fig.bitflip1}} \hfill
    \subfloat[]{\includegraphics[width=\columnwidth]{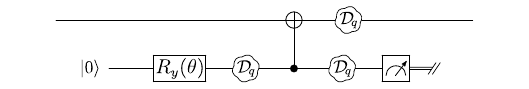} \label{fig.bitflip2}} \hfil
    \subfloat[]{\includegraphics[width=\columnwidth]{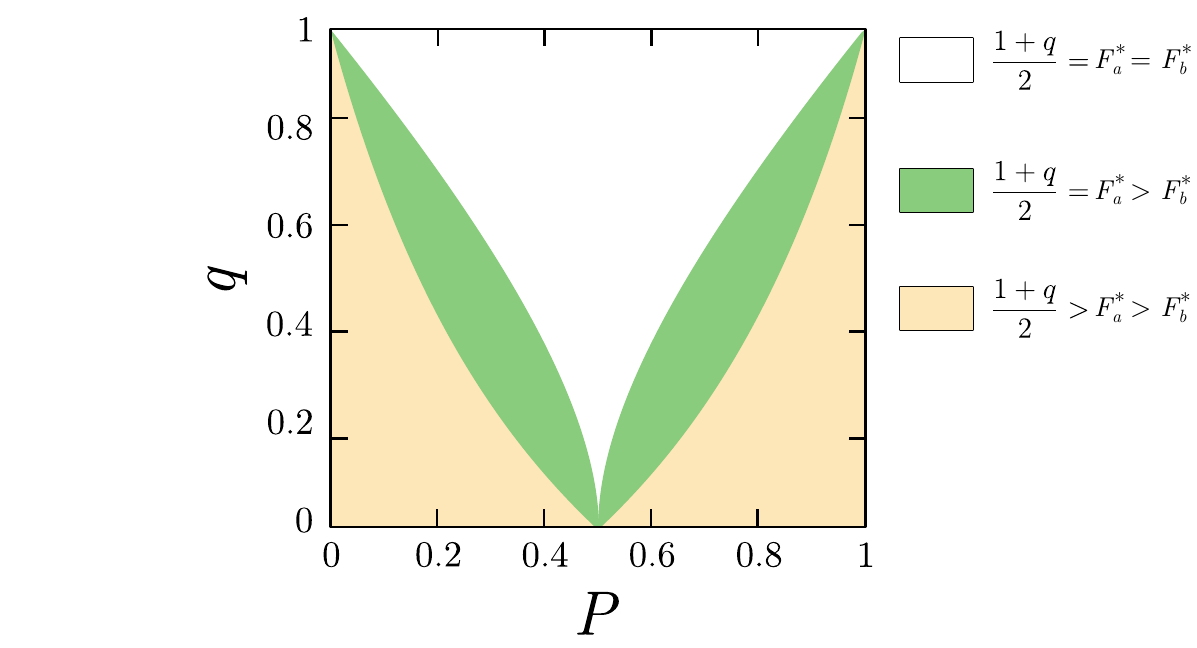} \label{fig:7c}}
    \caption{ \label{fig.7} (a,b) Two ways of implementing a bit flip channel. The channels are implemented in a quantum computer with (depolarizing) noisy gates. In (b) the angle $\theta$ is given by $\sin^2(\theta/2) = 1-p$, and $R_y = \text{exp}\{-i \frac{\theta}{2} Y\}$. (c) Comparison of the fidelity of the two circuits.}
\end{figure}

We briefly analyze the simulation of a bit-flip channel. We aim to implement the bit-flip channel given by
\begin{equation}
    \hat{\mathcal{E}}(\rho) = P \, \rho + (1-P) X \, \rho \, X.
\end{equation}
We consider the simulation by implementing the $X$ gate with probability $p$, see Fig.~\ref{fig.bitflip1}, and by using an ancillary system \ref{fig.bitflip2}. We assume that after each gate the system is affected by depolarizing noise, given by
\begin{equation}
    \hat{\mathcal{D}}_q(\rho) = q \rho + \frac{1-q}{2} \id.
\end{equation}
In this way, we find the Choi fidelity of each implementation is given by 
\begin{equation}
\begin{aligned}
    F_a & = \frac{1}{4} \left(\sqrt{P[(4 p-1)q+1]} + \sqrt{(1-P) [(3-4 p) q+1]} \right)^2 \\
    F_b & = \frac{1}{4} \left( \sqrt{P [(4 p-2) q^2 + q + 1]} \right.\\
    & \hspace{2.75cm} \left. + \sqrt{(1-P) [(2-4 p) q^2 + q + 1]} \right)^2,
\end{aligned}
\end{equation}
respectively. Note that $p$ is a parameter that we can freely tune. Therefore, for each value of $P$ and $q$ we can find the value of $p$ that maximizes $F^* = \max_p F$. In Fig. \ref{fig:7c} we show that for all values of $P$ and $q$ it is fulfilled that $F^*_b \leq F^*_a$. 

\section{Alternative circuits for amplitude damping channel}
\label{appendix.Alternative.circuits.for.amplitude.damping.channel}

\begin{figure}
    \centering
    \subfloat[]{\includegraphics[width=\columnwidth]{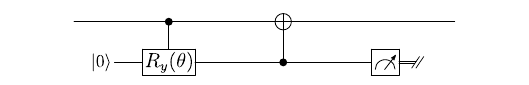}\label{fig.ampitude.damping.a}} \hfill
    \subfloat[]{\includegraphics[width=\columnwidth]{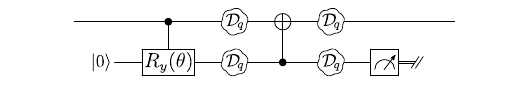}\label{fig.ampitude.damping.b}} \hfill
    \subfloat[]{\includegraphics[width=\columnwidth]{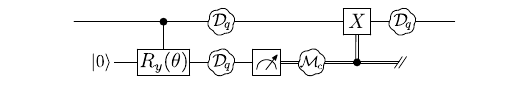}\label{fig.ampitude.damping.c}}
    \caption{\label{fig.ampitude.damping} (a) Ideal circuit to simulate amplitude damping noise. (b,c) The channels are implemented in a quantum computer with (depolarizing) noisy gates and noisy measurements. The angle is given by $\sin^2(\theta/2) = \gamma$.}
\end{figure}

The amplitude damping channel is given by Kraus operators,
\begin{align*}
    K_0 & = \proj{0} + \sqrt{1-\gamma} \proj{1}, \\
    K_1 & = \sqrt{\gamma} \ketbra{0}{1}.
\end{align*}
This channel (and in general any channel) can be simulated in a circuit fashion using different implementations, also assuming different computer noises. For instance, using an ancilla qubit system, the damping channel can be implemented as shown in Fig. \ref{fig.ampitude.damping.a}. In Fig. \ref{fig.ampitude.damping.b} we show how the circuit is affected by depolarizing noise after each gate, while in Fig. \ref{fig.ampitude.damping.c} we assume the same noise but the control-X gate is substituted by a measurement followed by a conditioned X gate.



\end{document}